\documentclass[prd,aps,twocolumn,superscriptaddress,nofootinbib,showpacs,10pt,preprintnumbers]{revtex4-1}

\usepackage{amsmath,amssymb,bm,multirow,float,bbold,hyperref}
\usepackage{graphicx}
\usepackage[american]{babel}

\usepackage{enumerate}
\newcommand{\lsim}{\stackrel{<}{_\sim}}
\newcommand{\gsim}{\stackrel{>}{_\sim}}

\newcommand{\be}{\begin{equation}}
\newcommand{\ee}{\end{equation}}

\newcommand{\sbt}{s_{\beta}}
\newcommand{\cbt}{c_{\beta}}

\usepackage{xcolor,colortbl}
\definecolor{Gray}{gray}{0.95}
\definecolor{RGray}{gray}{0.85}
\definecolor{CGray}{gray}{0.92}

\usepackage{color}

\newcommand{\beqn}{\begin{eqnarray}}

\newcommand{\bat}{\begin{array}{cc}}
\newcommand{\ea}{\end{array}}
\newcommand{\batt}{\begin{array}{ccc}}
\newcommand{\eeqn}{\end{eqnarray}}

\allowdisplaybreaks

\begin{document}

\title{
Family non-universal \texorpdfstring{\boldmath{$Z^{\prime}$}}{Z'} models with protected flavor-changing
interactions}

\author{Alejandro Celis}\thanks{alejandro.celis@physik.uni-muenchen.de}
 \affiliation{Ludwig-Maximilians-Universit\"at M\"unchen, 
   Fakult\"at f\"ur Physik,\\
   Arnold Sommerfeld Center for Theoretical Physics, 
   80333 M\"unchen, Germany}

\author{Javier Fuentes-Mart\'in}\thanks{javier.fuentes@ific.uv.es}
\affiliation{Instituto de F\'isica Corpuscular, Universitat de Val\`encia - CSIC, Apt. Correus 22085, E-46071 Val\`encia, Spain}

\author{Martin Jung}\thanks{martin.jung@tum.de}
\affiliation{
TUM Institute for Advanced Study, Lichtenbergstr. 2a, D-85747 Garching, Germany.}
\affiliation{
Excellence Cluster Universe, Technische Universit\"at M\"unchen, Boltzmannstr. 2, D-85748 Garching,
Germany.}

\author{Hugo Ser\^odio}\thanks{hserodio@kaist.ac.kr}
\affiliation{Department of Physics, Korea Advanced Institute of Science and Technology,\\
335 Gwahak-ro, Yuseong-gu, Daejeon 305-701, Republic of Korea   }

\date{\today}

\preprint{LMU-ASC 22/15}
\preprint{IFIC/15-23}
\preprint{FLAVOUR(267104)-ERC-95}

\begin{abstract}{
We define a new class of $Z^{\prime}$ models with neutral flavor-changing interactions at tree level
in the down-quark sector. They are related in an exact way to elements of the quark mixing matrix
due  to an underlying flavored $\mathrm{U(1)}^{\prime}$ gauge symmetry, rendering these models
particularly predictive. 
The same symmetry implies lepton-flavor non-universal couplings, fully determined by the gauge
structure of the model. Our models allow to address presently observed deviations from the SM
and specific correlations among the new physics contributions to the Wilson coefficients
$C_{9,10}^{(\prime) \ell}$ can be tested in $b \rightarrow s \ell^+ \ell^-$ transitions. We
furthermore predict lepton-universality violations in $Z^{\prime}$ decays, testable at the LHC.}
\end{abstract}


\maketitle

\section{Introduction}\label{s:intro}
The Standard Model (SM) of electroweak interactions~\cite{Glashow:1961tr}, based on the gauge
group $\mathrm{G}_{\mbox{\scriptsize{SM}}} \equiv \mathrm{SU(2)}_L \times \mathrm{U(1)}_Y$, has been
tested to very high accuracy over the last decades. This has strong implications for
new physics (NP) scenarios, which are then required to involve a very high mass scale, a
highly non-trivial flavor structure, or both. Recently the LHCb collaboration has performed
two measurements involving semileptonic $b\to s\ell^+\ell^-$
transitions which show deviations with respect to the SM expectations: 
The ratio $R_K = \mathrm{Br}(B \rightarrow K \mu^+ \mu^-)/\mathrm{Br}(B\rightarrow K e^+ e^-)$ has been measured with a central value indicating a
violation of lepton universality at the $25\%$ 
level~\cite{Aaij:2014ora}, implying a $2.6\sigma$ deviation, and the observable $P'_5$ obtained from the angular analysis of $B\to K^*\mu^+\mu^-$ decays differs from the SM expectation with $2.9\sigma$
significance~\cite{Aaij:2013qta}.
Taking these measurements at face value, they require the NP to have the following
features:
\begin{enumerate}[(i)]
  \item Sizable contributions to $b\to s\ell^+\ell^-$ transitions,
  \item lepton non-universal couplings.
\end{enumerate}

Extensions of the SM gauge group by an additional $\mathrm{U(1)}^\prime$ factor are among the
possible NP scenarios that could explain these
deviations~\cite{Descotes-Genon:2013wba,Buras:2013dea,Altmannshofer:2014cfa,Crivellin:2015mga,
Crivellin:2015lwa,Sierra:2015fma,Crivellin:2015era} 
(other NP interpretations have been discussed \emph{e.g.} in
Refs.~\cite{Hiller:2014yaa,Bhattacharya:2014wla,Varzielas:2015iva,Gripaios:2014tna}).
Such $\mathrm{U(1)}^\prime$ models have been popular extensions of the SM for many years, see
Ref.~\cite{Langacker:2008yv} and references therein. However, the majority of the extensions
considered in the literature are family universal in the quark (and charged-lepton) sector, and
therefore do not meet condition (ii). Models with departures from family universality can give rise
to large flavor changing $Z^{\prime}$ interactions, which are strongly constrained by current flavor
data~\cite{Langacker:2000ju,Crivellin:2015era,Buras:2012jb}.
The conditions (i) and (ii) therefore limit considerably the viable $\mathrm{U(1)}^{\prime}$ gauge
symmetries.

Starting a new construction with the minimal particle content, \emph{i.e.} the SM one without
right-handed neutrinos, the only possible extension is gauging one of the
combinations of family-specific lepton number, $L_\alpha-L_\beta$ (with $\alpha,\beta=e,\mu,\tau$
and $\alpha\neq \beta$)~\cite{He:1990pn}. Since the resulting $Z^\prime$ boson couples only to
leptons, these models do not meet requirement (i) and have been discussed mostly in the context of
neutrino phenomenology, usually considering the $L_\mu-L_\tau$
symmetry~\cite{Binetruy:1996cs}.
Recent models~\cite{Altmannshofer:2014cfa,Crivellin:2015mga} circumvent this problem by introducing
additional fermions (vector-like quarks). This yields the required couplings at an effective level.

In order to avoid such exotic vector-like quarks, the $\mathrm{U(1)}^\prime$ symmetry has to involve both
quarks and leptons, and should be family-dependent in order to satisfy conditions (i) and (ii).
However, such a non-trivial flavor symmetry in the quark sector requires the extension of the scalar
sector in order to accommodate the quark masses and mixing angles~\cite{Leurer:1992wg}. It has been
shown in Ref.~\cite{Crivellin:2015lwa} that this option can be realized by including an additional
complex scalar doublet. This model generates flavor-changing phenomena in the down-quark
sector which are related in a first approximation to the Cabibbo--Kobayashi--Maskawa (CKM) mixing
matrix~\cite{Cabibbo:1963yz}. Flavor-changing interactions are also present in the up-quark sector
of this model, but these are not related to CKM matrix elements or quark masses.

We would like our model to have \emph{all} of its fermion couplings related \emph{exactly} to
elements of the CKM matrix. To that aim, we note that there is a class of two-Higgs-doublet models
(2HDM) that achieves this, known as Branco--Grimus--Lavoura (BGL) models~\cite{Branco:1996bq}. These
models have flavor-diagonal interactions in the up-quark and charged-lepton sectors, together with
flavor-changing neutral currents (FCNCs) in the down-quark sector. The latter are suppressed by
quark masses and/or off-diagonal CKM elements in an exact way, thereby providing an alternative
solution to the flavor problem in 2HDMs which differs radically from the hypothesis of natural
flavor conservation~\cite{Glashow:1976nt}.

While in the original BGL model the flavor symmetry is global, in this work we promote it to a local
one. This is achieved by charging also the leptons under the symmetry, thereby
enabling anomaly cancellation. In this gauged BGL framework ($\mathrm{U(1)}^\prime_{\rm BGL}$), the
properties of the BGL models are transferred to the gauge boson sector: we obtain FCNCs mediated at
tree-level by the neutral scalar and massive gauge vector bosons of the theory, all of which are
suppressed by off-diagonal CKM elements and/or fermion masses, and therefore naturally suppressed.
This class of models necessarily exhibits deviations from lepton universality due to its gauge
structure. Note that this form of flavor suppression in the down-quark sector also appears in
certain 3-3-1 
models~\cite{Buras:2013dea}. 
However, in our framework this is not a
choice, but a consequence of the gauge symmetry.

This paper is organized as follows: We formulate the $\mathrm{U(1)}^\prime_{\rm BGL}$ models in
Sec.~\ref{s:frame}. In Sec.~\ref{sec:pheno} we discuss the constraints from flavor and collider
data, testing the strong correlations present in these models, specifically with respect to the
deviations observed by LHCb. We conclude in Sec.~\ref{sec:con}. The appendices include technical
details on anomaly cancellation and the scalar sector.

\section{Gauged BGL symmetry}\label{s:frame}  
A characteristic feature of BGL models~\cite{Branco:1996bq}\footnote{Their extension to
include neutrino masses has been discussed in Ref.~\cite{Botella:2011ne}.} is the presence of FCNCs
at tree level, which are however sufficiently suppressed by combinations of CKM matrix elements.
This is a consequence of specific patterns of Yukawa couplings, generated by corresponding charge
assignments under a horizontal, family-non-universal (BGL-)symmetry.

The quark Yukawa sector of the model reads ($i=1,2$)
\begin{align}\label{eq:Lagquarks}
-\mathcal{L}^{\mbox{\scriptsize{quark}}}_{\mbox{\scriptsize{Yuk}}}&=\overline{q_L^0}\, \Gamma_i\,
\Phi_i d^0_R+\overline{q_L^0}\, \Delta_i\, \widetilde{\Phi}_i  u^0_R+\text{h.c.} \,,
\end{align}         
where $\Gamma_i$ and $\Delta_i$ denote the Yukawa coupling matrices for the down- and up-quark sectors, respectively, and $\widetilde
\Phi_i \equiv i \sigma_2 \Phi_i^\ast$, with the Pauli matrix $\sigma_2$.
The neutral components of the Higgs doublets acquire vacuum expectation values (vevs) $|\langle
\Phi_i^0  \rangle  | = v_i/\sqrt{2}$. 
Defining
$v \equiv (v_1^2 + v_2^2)^{1/2}$, measurements of the muon lifetime fix $v= (\sqrt{2} G_F)^{-1/2}
\simeq 246$~GeV.
We further define $\tan \beta = v_2/v_1$ and use the following abbreviations:
$\cos \beta \equiv c_{\beta}$, $\sin \beta \equiv s_{\beta}$, $\tan \beta \equiv  t_{\beta}$.

Choosing an abelian symmetry under which a field $\psi$ transforms as
\begin{equation}
\psi\rightarrow e^{i\mathcal{X}^{\psi}}\psi\,,
\end{equation}
the most general symmetry transformations in the quark sector yielding the required textures are of the form
\begin{align}
\begin{aligned}  \label{eq:system}
\mathcal{X}_L^q&=\frac{1}{2}\left[\mathrm{diag}\left(X_{uR},X_{uR},X_{tR}\right)+X_{dR}\,\mathbb{1}\right]\,,\\
\mathcal{X}_R^u&=\mathrm{diag}\left(X_{uR},X_{uR},X_{tR}\right)\,,\\
\mathcal{X}_R^d&=X_{dR}\,\mathbb{1}\,,
\end{aligned}
\end{align}
with $X_{uR}\neq X_{tR}$. 
The Higgs doublets transform as
\begin{align}
\begin{aligned}
\mathcal{X}^\Phi&= \mathrm{diag}\,(X_{\Phi_1}, X_{\Phi_2})\\
&= \frac{1}{2}\mathrm{diag}\,\left(X_{uR}-X_{dR},X_{tR}-X_{dR}\right)\,.
\end{aligned}
\end{align}
There are several possible implementations of this symmetry, related by permutations in flavor
space and exchanging up- and down-quark sectors; here,
for definiteness, the top quark has been singled out, yielding the patterns
%
%%%%%
\beqn \label{eq:BGLText}
\begin{aligned}
\Gamma_1&=
\begin{pmatrix}
\ast&\ast&\ast\\
\ast&\ast&\ast\\
0&0&0
\end{pmatrix}\,,\quad
\Gamma_2=
\begin{pmatrix}
0 & 0 & 0\\
0 & 0 & 0\\
\ast&\ast&\ast
\end{pmatrix}\,, \\[0.2cm]
\Delta_1&=
\begin{pmatrix}
\ast & \ast&0\\
\ast& \ast&0\\
0 & 0 &0
\end{pmatrix} \,\,,\quad
\Delta_2=
\begin{pmatrix}
0 &0 &0\\
0 & 0&0\\
0 &0 &\ast
\end{pmatrix} \,. 
\end{aligned}
\eeqn
%%%%%
These textures give rise to FCNCs in the down-quark sector, 
which are however suppressed by quark masses and off-diagonal elements of the third CKM
row~\cite{Branco:1996bq}.
This is related to the fact that the CKM mixing matrix $V\equiv U_{uL}^\dagger
U_{dL}$ has a direct correlation in this model with the elements of the third row of the
rotation of the left-handed down quarks, \emph{i.e.}
$\left(U_{dL}\right)_{3i}=V_{3i}$. The choice of the top quark in Eq.~\eqref{eq:system} implies
a particularly strong suppression of flavor-changing phenomena in light-quark systems. Present
constraints from $\Delta F=1$ and $\Delta F=2$ quark flavor transitions are accommodated even when
the scalars of the theory are light, with masses of $\mathcal{O}(10^2)$~GeV~\cite{Botella:2014ska}.

While this model can be implemented through an abelian discrete symmetry, this always yields
an accidental continuous abelian symmetry which introduces an undesired
Goldstone boson into the theory~\cite{Branco:1996bq}. Several solutions to this problem have
already been proposed in the literature~\cite{Branco:1996bq,Celis:2014iua}; in the following we 
provide a new solution, promoting the BGL symmetry to a local one.

When gauging a symmetry, special attention must be paid to whether it remains
anomaly-free.\footnote{In this work we consider only symmetries as ``good'' if all gauge anomalies are canceled. It would be also possible to consider anomalous $\mathrm{U(1)}$ extensions in the
low-energy theory whose anomaly graphs get canceled via the so-called Green-Schwarz
mechanism~\cite{Green:1984sg}.} 
BGL models are automatically free of QCD anomalies, i.e.
$\mathrm{U(1)}^\prime[\mathrm{SU(3)}_C]^2$~\cite{Celis:2014iua}. However, 
we also need to fulfill the anomaly conditions for the following combinations:
\begin{align}\label{eq:anomalies}
&\mathrm{U(1)}^\prime [\mathrm{SU(2)}_L]^2\,,\quad \mathrm{U(1)}^\prime [\mathrm{U(1)}_Y]^2\,, 
\qquad [\mathrm{U(1)}^\prime]^2 \mathrm{U(1)}_Y \,,\nonumber\\
&[\mathrm{U(1)}^\prime]^3\,,\quad  \qquad \quad \,  \mathrm{U(1)}^\prime [\mbox{Gravity}]^2 \,.
\end{align}
We find that there is no solution for this system within the quark sector alone  with the charge
assignments in Eq.~\eqref{eq:system}. Satisfying these anomaly conditions is highly non-trivial and
requires, in general, additional fermions. However, the implementation of the BGL symmetry as a
local symmetry is possible by adding only the SM leptonic sector when allowing for lepton-flavor
non-universal couplings. As in the SM, the cancellation of the gauge anomalies then occurs due to a
cancellation between quark and lepton contributions; the cancellation  will however involve all
three fermion generations, contrary to the SM gauge group, for which the anomaly cancellation occurs
separately within each fermion generation.

Gauging the BGL symmetry therefore not only removes the unwanted Goldstone boson, which is ``eaten'' 
by the $Z^\prime$, but also provides a consistent $Z^\prime $ model without extending the
particle content beyond the additional scalars, \emph{i.e.} no vector-like quarks are necessary.
The flavor structure is extended to the $Z^\prime $ couplings, yielding again strongly suppressed FCNCs
in the down-type quark sector, while the up-type-quark couplings remain diagonal.
Especially, since the flavored gauge symmetry fixes also the charged-lepton Yukawa matrices to
be diagonal, we can obtain large deviations from lepton universality without introducing
lepton-flavor violation. Our models provide thereby explicit examples for which the general
arguments given in Refs.~\cite{Glashow:2014iga,Bhattacharya:2014wla,Boucenna:2015raa} do not hold.

\subsection{Lepton sector} \label{sec:leptons}

As emphasized above, the anomaly constraints can be fulfilled
by assuming that leptons are also charged under the $\mathrm{U(1)}^{\prime}$ symmetry. The
resulting charge constraints
are given in Appendix~\ref{sec:anomaly}. 

For $X_{\Phi_2}=0$ the mixing between the neutral massive gauge bosons is suppressed for
large $\tan\beta$. With this choice, we obtain only one possible solution to the anomaly conditions
up to lepton-flavor permutations, implying the following charge assignments:
\begin{align}  \label{eq:charges_model}
\mathcal{X}_R^d&=\mathbb{1}\,,\quad
&\mathcal{X}_R^u&=\text{diag}\left(-\frac{7}{2},-\frac{7}{2},1\right)\,,\nonumber\\
\mathcal{X}_L^q&=\text{diag}\left(-\frac{5}{4},-\frac{5}{4},1\right)\,,\quad
&\mathcal{X}_L^\ell&=\text{diag}\left(\frac{9}{4},\frac{21}{4},-3\right)\,,\nonumber\\
\mathcal{X}_R^{e}&=\text{diag}\left(\frac{9}{2},\frac{15}{2},-3\right)\,,\quad
&\mathcal{X}^{\Phi}&=\text{diag}\left(-\frac{9}{4},0\right)\,.
\end{align}
We have normalized all 
$\mathrm{U(1)}^{\prime}$ charges by setting $X_{dR}=1$, thereby fixing also the
normalization for the gauge coupling $g^\prime$.
In addition to the top, this
particular model 
singles out the tau lepton as the only one coupling to $\Phi_2$.
Permutations of the charges in $\mathcal{X}_L^\ell$ and $\mathcal{X}_R^{e}$ 
give rise to 5 more models.

With the charge transformations in Eq.~\eqref{eq:charges_model}, the charged-lepton Yukawa
sector takes the form
\begin{align}
-\mathcal{L}^{\mbox{\scriptsize c-leptons}}_{\mbox{\scriptsize{Yuk}}}&=\overline{\ell_L^0}\, \Pi_i\, \Phi_i e^0_R+\text{h.c.} \,,
\end{align}
with
\begin{equation}
\Pi_1=
\begin{pmatrix}
\ast&0&0\\
0&\ast&0\\
0&0&0
\end{pmatrix}\,,\quad \Pi_2=
\begin{pmatrix}
0&0&0\\
0&0&0\\
0&0&\ast
\end{pmatrix}\,,
\end{equation}
while the quark Yukawa sector remains unchanged.

We call to attention that no reference to the neutrino sector has been made in the previous
arguments. In this work we are mostly interested in the properties of the quark and charged-lepton
sectors in the presence of a new gauge boson, while we leave a detailed discussion of the
neutrino sector to future work.
For a possible implementation with 
the SM fermion
content we can build the $d=5$ Weinberg effective operator
$(\overline{\ell}\tilde{\Phi}_a)(\tilde{\Phi}_b^T\ell^c)$, which after spontaneous symmetry breaking
induces a Majorana mass term for the neutrinos. If this effective operator is invariant under the
new flavored gauge symmetry in Eq.~\eqref{eq:charges_model}, we are led to a $L_\mu-L_\tau$
symmetric $d=5$ effective operator and, consequently, Majorana mass term. This structure has been
studied in Refs.~\cite{Binetruy:1996cs},
and while it is not capable of fully accommodating the neutrino data, it can serve as a good
starting point.

If we want to improve on this scenario while keeping our solution to the anomaly conditions,
we may want a mechanism that at the effective level already breaks the accidental
$L_\mu-L_\tau$ symmetry.
For instance, we may extend the model with 3 right-handed neutrinos, where one of them is not
charged while the other two have opposite charges. In this way the anomaly solutions remain
unchanged. Choosing the charges of the right-handed neutrinos appropriately,
the Dirac mass term for the neutrinos can be made diagonal and the Majorana mass for the
right-handed ones $L_\mu-L_\tau$ symmetric. Along the lines of
Ref.~\cite{Crivellin:2015lwa}, coupling an additional scalar to the right-handed neutrinos could
break the accidental symmetry at the effective level. 

We can also envisage other mechanisms that do change the anomaly conditions and, therefore,
introduce new solutions.

\subsection{Gauge boson sector} \label{sec:gaugesector}
In order to avoid experimental constraints on a $Z^\prime$ boson, the $\mathrm{U(1)}^{\prime}$ symmetry
must be broken at a relatively high scale, rendering the new gauge boson significantly heavier
than the SM ones~\cite{Langacker:2008yv}. This can be achieved through the introduction of a complex
scalar singlet $S$, charged under the BGL symmetry with charge $X_S$, which acquires a vev
$|\langle S \rangle | = v_S/\sqrt{2}\gg v$.
The charge of the scalar singlet is fixed once the scalar potential is specified, in our case $X_S
=-9/8$, see Appendix~\ref{app:scalar} for details.

As a consequence of the gauge symmetry breaking, $\mathrm{G}_{\mbox{\scriptsize{SM}}} \times 
\mathrm{U(1)}^{\prime}  \rightarrow \mathrm{U(1)}_{\mbox{\scriptsize{em}}}$, the neutral massive
gauge bosons mix, giving rise to a Lagrangian of the form\footnote{We neglect here a possible
mixing coming from the gauge kinetic Lagrangian as the mass mixing shows already all relevant
qualitative effects~\cite{Langacker:2008yv}.}
\begin{align}
\begin{aligned}
\mathcal{L}=&\,-\frac{1}{4}A_{\mu\nu}A^{\mu\nu}-\frac{1}{4}\hat{Z}_{\mu\nu}\hat{Z}^{\mu\nu}-\frac{1}{4}\hat{Z}^\prime_{\mu\nu}\hat{Z}^{\prime\,\mu\nu}   \\
&+\frac{1}{2}\hat{M}_{Z}^2\hat{Z}_{\mu} \hat{Z}^\mu+\frac{1}{2}\hat{M}_{Z^\prime}^2\hat{Z}^\prime_{\mu}\hat{Z}^{\prime\,\mu}+\Delta^2 \hat{Z}^\prime_\mu \hat{Z}^{\mu}    \\
&-J_A^{\,\mu} A_\mu-\hat{J}_Z^{\,\mu} \hat{Z}_\mu-\hat{J}_{Z^\prime}^{\,\mu} \hat{Z}^\prime_\mu\,,
\end{aligned}
\end{align}
where $J_A^{\,\mu}$ and $\hat{J}_Z^{\,\mu}$ are the SM currents of the photon and the $Z$,
respectively.  The fermionic piece of the new current takes the form
\begin{align}
\hat{J}_{Z^\prime}^{\,\mu}\supset g^\prime\,\overline{\psi_{i}}\gamma^\mu \left[\left(\mathcal{ \widetilde X}^\psi_{L}\right)_{ij}P_L+\left(\mathcal{ \widetilde X}^\psi_{R}\right)_{ij}P_R\right] \psi_{j}\,,
\end{align}
with $P_{L,R} = (1\mp\gamma_5)/2$ denoting the usual chiral projectors, $g^\prime$ the
$\mathrm{U(1)}^\prime$ gauge coupling, and $\mathcal{\widetilde X}^\psi_{X}$ the phase
transformation matrices in the fermion mass basis, \emph{i.e.}
\begin{align}
\begin{aligned}
\mathcal{\widetilde X}^\psi_{R} &= \mathcal{X}_{R}^{\psi}\,,\quad  
\mathcal{\widetilde X}^u_{L} =  \mathcal{X}_{L}^{q} \,, \quad \mathcal{\widetilde X}^e_{L} = 
\mathcal{X}_{L}^{\ell}
\,,\quad{\rm and}\\
\label{eq:flavorstr}
  \mathcal{\widetilde X}^d_{L} &=  - \frac{5}{4} \mathbb{1} +\frac{9}{4}   \, \begin{pmatrix}
 |V_{td}|^2  &    V_{ts} V_{td}^*   &    V_{tb}  V_{td}^*      \\
  V_{td}  V_{ts}^*     &    |V_{ts}|^2   &   V_{tb} V_{ts}^*   \\
   V_{td}  V_{tb}^*  &     V_{ts}  V_{tb}^*&     |V_{tb}|^2 
\end{pmatrix}       \,.
\end{aligned}
\end{align}

The neutral gauge boson mass and mass-mixing parameters
in this model are given by
\begin{align}
\begin{aligned}
\hat{M}_Z^2&=\frac{e^2v^2}{4\hat{s}_W^2\hat{c}_W^2}\,,\\
\hat{M}_{Z^\prime}^2&=g^{\prime\,2}\left(v_i^2 X_{\Phi_i}^2+ v_S^2 X_S^2\right)\,,\\
\Delta^2&=-\frac{e\,  {g}^\prime}{2\hat{c}_W\hat{s}_W}X_{\Phi_i}v_i^2\,,
\end{aligned}
\end{align}
where $e=\sqrt{4\pi\alpha}$ is the electromagnetic charge and its ratio
with the $\mathrm{SU(2)}_L$ coupling $g$ defines $\hat{\theta}_W\equiv \arcsin\,(e/g)$, which differs
from the experimentally measured weak angle due to mixing effects~\cite{Babu:1997st}:
$\hat{s}_W\hat{c}_W\hat{M}_Z=s_Wc_W M_Z$, where $M_Z$ is the physical $Z$ mass.   We rotate to the
physical eigenbasis by performing the following orthogonal transformation:
\begin{align}
\begin{pmatrix}
Z_\mu\\
Z^{\prime}_\mu
\end{pmatrix} &= O
\begin{pmatrix}
\hat{Z}_\mu\\
\hat{Z}_\mu^\prime
\end{pmatrix}\,,\quad\text{with}\quad
O=
\begin{pmatrix}
c_\xi&s_\xi\\
-s_\xi&c_\xi
\end{pmatrix} \quad {\rm and}\nonumber\\
\tan 2\xi&=\frac{2\Delta^2}{\hat{M}_Z^2-\hat{M}_{Z^\prime}^2}\,.
\end{align}
Since we assume $v\ll v_S$, we obtain $\hat{M}_Z^2,\, \Delta^2 \ll\hat{M}_{Z^\prime}^2$ and
the resulting mixing angle is small~\cite{Langacker:2008yv}:
\begin{align}
\begin{split}
g^\prime \xi&\simeq\frac{e}{2 \hat{c}_W\hat{s}_W}\frac{X_{\Phi_1} v_1^2+X_{\Phi_2} v_2^2}{X_S^2 v_S^2} \simeq -\frac{9e}{8 \hat{c}_W\hat{s}_W}   \left(   \frac{ g^{\prime}   c_{\beta}   v  }{   M_{Z^{\prime}}  }      \right)^2  .
\end{split}
\end{align}%

Expanding in this small parameter, we obtain the following leading expression for
the hatted weak angle in terms of physical quantities
\begin{align}
\hat s_W^2 &\simeq s_\ast^2\equiv
s_W^2-\frac{c_W^2s^2_W}{c_W^2-s_W^2}\xi^2\left(\frac{M_{Z^\prime}^2}{M_{Z}^2}-1\right) \,.
\end{align}

In the physical basis the neutral gauge boson masses read
\begin{align}
\begin{aligned}
M_{Z,\,Z^\prime}^2=\frac{1}{2}\left[\hat{M}_{Z^\prime}^2+\hat{M}_Z^2\mp\sqrt{\left(\hat{M}_{Z^\prime}^2-\hat{M}_Z^2\right)^2+4\Delta^4} \,\right]\,,
\end{aligned}
\end{align}
and their fermionic currents are given by
\begin{align}
J_{Z^{(\prime)}}^\mu&\supset\overline{\psi_{i}}\gamma^\mu \left[\epsilon_{L,ij}^{(\prime) \psi}P_L+\epsilon_{R,ij}^{(\prime) \psi}P_R\right] \psi_{j}\,,
\end{align}%
with the couplings (up to $\mathcal{O}\left(\xi^2\right)$)
\begin{align}
\begin{aligned}
\epsilon^\psi_{X,ij}=&\,\frac{e}{c_Ws_W}\left[1+\frac{\xi^2}{2}\left(\frac{M_{Z^\prime}^2}{M_{Z}^2}-1\right)\right]\\
&\times\left[\left(T_3^{\psi_X}-s_*^2Q_\psi\right)\mathbb{1}+\xi\frac{g^\prime c_Ws_W}{e}\mathcal{ \widetilde X}^\psi_{X}\right]_{ij}\,,\\
\epsilon^{\prime\,\psi}_{X,ij}=&\,\frac{e}{c_Ws_W}\left[\frac{g^\prime c_Ws_W}{e}\mathcal{ \widetilde
X}^\psi_{X}-\xi\left(T_3^{\psi_X}-s_W^2Q_\psi\right)\mathbb{1}\right]_{ij}\,.
\end{aligned}
\end{align}
Here $T_3^{\psi_L}$ is the third component of weak isospin for the left-handed fields
($T_3^{\psi_R}=0$) and $Q_\psi$ the electric charge. 
We also define the vector and axial-vector combinations of the $Z^{(\prime)}$ couplings,
\be
\epsilon^{(\prime)\,\psi}_{V,ij} \equiv \epsilon^{(\prime)\,\psi}_{L,ij} + \epsilon^{(\prime)\,\psi}_{R,ij} \,, \qquad  \epsilon^{(\prime)\,\psi}_{A,ij} \equiv \epsilon^{(\prime)\,\psi}_{R,ij} - \epsilon^{(\prime)\,\psi}_{L,ij} \,.
\ee
The coefficients $\epsilon^{\psi}_{V (A),ij}$ encode corrections to the SM $Z$ couplings due to
mixing with the $Z^\prime$ which are proportional to $g^{\prime} \xi$ at leading order in the
mixing angle.
Such corrections are restricted to be small $( g^{\prime} \xi \lesssim
10^{-4})$~\cite{Agashe:2014kda,Erler:1999ub}.  In addition to our initial assumption,
$v_S \gg v$, which already gives a small $\xi$, we will assume that the mixing angle receives an
additional suppression, because $\tan {\beta}$ is large.\footnote{This limit is quite natural in our
model since $\Phi_2$ is coupled to the top quark, while $\Phi_1$ is coupled to the light first two
generations. Therefore, the hierarchy $v_2> v_1$ accommodates well the quark mass spectrum.} This
also guarantees that flavor-changing effects mediated by the $Z$ are negligible compared with those
of the $Z^{\prime}$.

Another important theoretical constraint is obtained when requiring the absence of a low-energy
Landau pole, \emph{i.e.} an energy scale $\Lambda_{\mbox{\scriptsize{LP}}}$ at which perturbativity
is lost. This scale can be found from the renormalization-group running of the $\mathrm{U(1)}^\prime$ coupling, 
we have:
\begin{equation}
\frac{d\alpha^\prime}{d\ln q^2}=b\,\alpha^{\prime \,2}+\mathcal{O}(\alpha^{\prime\, 3})\,,
\end{equation}
where the one-loop beta function $b$
contains the charge information of the model. It reads
\begin{equation}
b=\frac{1}{4\pi}\left[\frac{2}{3}\sum_{f}X_{fL,R}^2+\frac{1}{3}\left(2\sum_iX_{\Phi i}^2+X_S^2\right)\right]\,,
\end{equation}
with $f$ including all fermion degrees of freedom and $i=1,2$.
The Landau pole can be extracted from the pole of
$\alpha^{\prime}(q^2) =g^{\prime\,2}(q^2)/4\pi$, we obtain
\begin{equation}
\Lambda_{\mbox{\scriptsize{LP}}}\simeq M_{Z^\prime}\,
\text{exp}\left[\frac{1}{2\,b\,\alpha^\prime(M_{Z^\prime}^2)}\right]\,.
\end{equation}

For the charges in our model we have $b\simeq12.90$, see Eq.~\eqref{eq:charges_model}. Taking
$M_{Z^\prime}\simeq 4$~TeV, we get a Landau pole at the Planck scale,
$\Lambda_{\mbox{\scriptsize{LP}}}\gsim 10^{19}$~GeV, for $g^\prime\lsim0.12$. Notice that this bound
is stronger than the na\"ive perturbativity bound $\alpha^\prime(M_{Z^\prime}^2)\lsim
\mathcal{O}(1/\max\{|X_i|\}^2)$, which in our model gives $g^\prime \lsim 0.47$. We can relax this
condition by assuming that additional NP will appear at the see-saw or the Grand
Unification scale. This way we obtain $g^\prime\lsim0.14$ for $\Lambda_{\mbox{\scriptsize{LP}}}\geq10^{14}$~GeV
and $g^\prime\lsim0.13$ for $\Lambda_{\mbox{\scriptsize{LP}}}\geq10^{16}$~GeV. The extension of the
model to these scales is beyond the scope of this work.

\subsection{Scalar sector} \label{sec:scalarsec}
Here we discuss the main features of the Higgs sector of the model, a detailed derivation of these
results is given in Appendix~\ref{app:scalar}. It %
consists of two complex Higgs doublets and a complex scalar gauge singlet. Their properties are
largely determined by the fact that $v/v_S\ll 1$. The spectrum %
contains four would-be Goldstone bosons giving mass to the electroweak (EW) gauge vector bosons
$\{M_W, M_Z, M_{Z^{\prime}}\}$, a charged Higgs $H^{\pm}$ and four neutral scalars (three CP-even
$\{H_1, H_2, H_3\}$ and a CP-odd $A$).

We obtain a decoupling scenario with a light SM-like Higgs boson, $M_{H_1}^2  \sim
\mathcal{O}(v^2)$, three heavy quasi-degenerate states coming mainly from the scalar doublets
$M_{H_2}^2 \simeq  M_{A}^2 \simeq  M_{H^{\pm}}^2 \sim \mathcal{O}(v_S^2)$ and another heavy CP-even
Higgs coming mainly from the scalar singlet $M_{H_3}^2  \sim \mathcal{O}(v_S^2)$.\footnote{It is also possible to motivate scenarios that avoid decoupling based on an enhanced Poincar\'e symmetry protecting the weak scale~\cite{Foot:2013hna}. These can give rise to a rich scalar sector at
the EW scale and a pseudo-Goldstone boson in the spectrum.} The Yukawa
interactions of the scalar bosons reflect this decoupling.
They %
can be written as
\begin{align} \label{YukaLag}
-\mathcal{L}_Y&\supset  \sum_{\varphi = H_k, A}  \varphi \,  \left[ \overline{d_L}  \, Y_{d}^{\varphi}  \, d_R  +   \overline{u_L}  \, Y_{u}^{\varphi}  \, u_R   + \overline{\ell_L}  \, Y_{\ell}^{\varphi}  \, e_R   \right]  \nonumber \\
&+\frac{\sqrt{2}}{v }H^+\left(\overline{u_L}N_dd_R-\overline{u_R}N_u^\dagger d_L    +  \overline{\ell_L}N_{\ell}e_R  \right)+\text{h.c.}
\end{align}
Neglecting corrections of $\mathcal{O}(v^2/v_S^2)$ we obtain
\begin{align}
\begin{aligned}
 Y_{f}^{H_1}  &= \frac{D_{f}}{v}   \,, \qquad   \qquad ~~~  Y_{f}^{H_2} = - \frac{N_{f}}{v}    \,,  \\
 Y_{u}^{A} &=  i \frac{N_{u}}{v} \,, \qquad    \qquad \, \, \, \, \, \,  Y_{d,\ell}^{A} = - i \frac{N_{d,\ell}}{v} \,,
 \end{aligned}
 \end{align}
with ($f=u,d,\ell)$. Here fermions have been rotated to the mass eigenbasis. The
diagonal matrices $D_{f=u,d,\ell}$ contain
the fermion masses, and the matrices $N_{f}$ are given as
\begin{align}\label{eq:YBGL}
\left(N_{d}\right)_{ij}\;=\;&\frac{v_2}{v_1}\left(D_d\right)_{ij}-\left(\frac{v_2}{v_1}+\frac{v_1}{v_2}\right)(V^\dagger)_{i3}(V)_{3j}(D_d)_{jj}
\,,  \nonumber\\
N_u\;=\;& \frac{v_2}{v_1}\text{diag}(m_u,m_c,0) -\frac{v_1}{v_2}\text{diag}(0,0,m_t)  \,,
\nonumber\\
N_{\ell}\;=\;& \frac{v_2}{v_1}\text{diag}(m_e,m_{\mu},0) -\frac{v_1}{v_2}\text{diag}(0,0,m_{\tau})  \,.
\end{align}
The couplings of the scalar $H_3$ to the fermions are additionally suppressed,
since they are generated only through mixing with the doublet degrees of freedom: $Y_{f}^{H_3}
\simeq \mathcal{O}(v/v_S)$.
The matrix $N_d$ is non-diagonal in flavor space, giving again rise to tree-level FCNCs in the
down-quark sector controlled by the CKM matrix $V$. Furthermore, as shown in
Appendix~\ref{app:scalar}, the constraints from the $\mathrm{U(1)}^\prime$ symmetry render
the scalar potential CP invariant.
CP violation in our model is therefore uniquely controlled by the CKM phase, like in the
SM.\footnote{We do not consider strong CP violation in this work.}

\subsection{Model variations} \label{sec:variations}

As mentioned in section \ref{sec:leptons}, we have six model implementations according to the
different lepton flavor permutations of the $\mathrm{U(1)}^\prime$ charges, with identical
quark and scalar charges given in Eq.~\eqref{eq:charges_model}.
These permutations are performed in the basis where the lepton mass matrix is diagonal and
the eigenvalues are correctly ordered. This way each permutation corresponds to the simultaneous
interchange of the charges of both left- and right-handed leptons. It is straightforward to
check that the permutation of just one set of charges, while satisfying the anomaly conditions
as well, would give rise to a non-diagonal mass matrix and reduce to one of the six cases
considered here after diagonalization.

In order to present the predictions for each of these models, we introduce the generic lepton
charges
\begin{align}
\begin{aligned}
X_{1L}&=\frac{9}{4}\,,\qquad X_{2L}=\frac{21}{4}\,,\qquad X_{3L}=-3\,,\\
X_{1R}&=\frac{9}{2}\,,\qquad X_{2R}=\frac{15}{2}\,,\qquad X_{3R}=-3\,,
\end{aligned}
\end{align}
such that each implementation is labeled by $(e,\,\mu,\,\tau)=(i,\,j,\,k)$. For instance,
the model presented in Eq.~\eqref{eq:charges_model} is now labeled as $(1,\,2,\,3)$.

The additional possibilities to implement the quark and neutrino sectors have been discussed in
Secs.~\ref{s:frame} and~\ref{sec:leptons}, respectively.

\section{Discussion} \label{sec:pheno}
In this section we discuss the phenomenology of the gauged BGL models introduced in the previous
section. The phenomenology of a scalar sector with a flavor structure identical to the one
of our model has been analyzed in Ref.~\cite{Botella:2014ska}. Since we are assuming a
decoupling scalar sector, we can naturally accommodate a SM-like Higgs at $125$~GeV; the heavy
scalars are not expected to yield sizable contributions to flavor observables in general. We
therefore focus on the phenomenological implications of the $Z^{\prime}$ boson for this class of
models. In the cases where the constraints depend on the choice of lepton charges in our model, we
give here the most conservative bound and discuss the differences in Secs.~\ref{sec::models}
and~\ref{npsec}.

\subsection{Low-energy constraints}

Despite the large mass of several TeV, the $Z^\prime$ boson yields potentially significant
contributions on flavor observables, due to its flavor-violating couplings in the down-type quark
sector. However, because of $M_W^2/M_{Z'}^2\leq 0.1\%$
and the CKM suppression of the flavor-changing $Z'$ couplings, these contributions can only be
relevant when the corresponding SM amplitude has a strong suppression in addition to $G_F$. This is
specifically the case for meson mixing amplitudes and electroweak penguin processes which we will
discuss below. Further examples are differences of observables that are small in the SM, for
example due to lepton universality or isospin symmetry.

The particular
flavor structure of the model implies a strong hierarchy for the size of different flavor
transitions:  $|V_{ts}^* V_{td}| \sim \lambda^5 \ll |V_{td}^* V_{tb}| \sim \lambda^3 \ll  |V_{ts}^*
V_{tb}| \sim \lambda^2$ (where $\lambda \simeq 0.226$). However, since the SM amplitudes often
show a similar hierarchy, the relative size of the $Z'$ contribution has to be determined on a
case-by-case basis. We obtain for example similar bounds from the mass differences $\Delta
m_{d,s}$ in the $B_{d,s}$ systems and $\epsilon_K$. Here we include exemplarily the constraint from
$B_s$ mixing, while leaving a detailed phenomenological analysis for future work.

Given the high experimental precision of $\Delta m_s$~\cite{Amhis:2014hma}, the strength of the
corresponding constraint depends completely on our capability to predict the SM value. The limiting
factors here are our knowledge of the hadronic quantity $f_{B_s}\sqrt{\hat B_s}$ (for which
substantial progress is expected soon, see Ref.~\cite{LatticeProgress}) and the relevant CKM elements.
Using the corresponding values from Refs.~\cite{Aoki:2013ldr} and~\cite{Charles:2004jd},
respectively, we conclude that contributions to $\Delta m_s$ up to $20\%$ remain possible at $95\%$~confidence level (CL);
from this we obtain the limit $M_{Z^\prime}/ g^{\prime}\gtrsim 16~\text{TeV}$.
Tree-level scalar contributions 
have been neglected, since their effect cancels to a very good approximation in the decoupling
limit considered here~\cite{Buras:2013uqa}. 

Regarding the input for the CKM parameters, we make the following observations in our models: 
(i) Charged-current tree-level processes receive negligible contributions. (ii) The mixing
phase in $B_{d,s}$ mixing and the ratio $\Delta m_s/\Delta m_d$ remain SM-like.
(iii) The additional direct CP violation in $B\to \pi\pi,\rho\pi,\rho\rho$ and $B\to J/\Psi K$ is
negligible. These observations imply that in our context the global fits from Refs.
\cite{Charles:2004jd,Ciuchini:2000de} remain valid to good approximation.

In our models the $Z^{\prime}$ boson couples to muons and will contribute to neutrino trident
production (NTP), $\nu_{\mu} N \rightarrow \nu N \mu^+ \mu^-$.  We expect dominance of the
$Z^{\prime}$ vector current for our models. The NTP cross section normalized to the
SM one was calculated in Refs.~\cite{Altmannshofer:2014cfa,Altmannshofer:2014pba,Crivellin:2015era}. Using a
combination of the latest NTP cross-section measurements we obtain a bound on the
parameter combination $M_{Z^{\prime}}/g^{\prime}$~\cite{Geiregat:1990gz},
which is however weaker than the one
from $B_s$ mixing. 

Atomic parity violation (APV) measurements also place bounds on additional neutral gauge bosons coupling to electrons and light quarks.   In particular, the precise measurement of the $^{\mbox{\scriptsize{133}}}\text{Cs}$ weak charge from APV experiments is in a reasonable agreement with the prediction
of the SM and can be used to set bounds on the $Z^{\prime}$ boson of our
models~\cite{Wood:1997zq}.
The bound we obtain on $M_{Z^{\prime}}/g^{\prime}$ from the latest determination of the
$^{\mbox{\scriptsize{133}}}\text{Cs}$ weak charge is however again weaker than the limit from $B_s$
mixing.

Further potentially strong bounds stem from electric dipole moments (EDMs) and the anomalous
magnetic moment of the muon. EDMs from the scalar sector occur at the two-loop level; while this
alone does not render them necessarily sufficiently small, the additional suppression and
cancellations in the decoupling limit imply very small
effects~\cite{Botella:2012ab,Botella:2014ska}.
Anomalous magnetic moments can be present at the one loop level, but are again too small
in 2HDMs near the decoupling limit~\cite{Crivellin:2013wna,Botella:2014ska}. 
Concerning the new gauge boson, one-loop contributions to EDMs require
different phases for left- and right-handed $Z^{\prime}$ couplings to quarks and/or
leptons~\cite{Chiang:2006we}; this feature is not present in our model, therefore the $Z^\prime$
contributions are again at the two-loop level and sufficiently suppressed, given the large
$Z^\prime$ mass.
In the case of anomalous magnetic moments we have non-vanishing one-loop contributions, which for
the muon read~\cite{Jegerlehner:2009ry}
\begin{equation}
a_\mu^{\mbox{\scriptsize{NP}}} \simeq \frac{m_\mu^2}{4\pi^2}\frac{g^{\prime
2}}{M_{Z^\prime}^2}\left(\frac{1}{3}X_{\mu V}^2-\frac{5}{3}X_{\mu A}^2\right)\,,
\end{equation} 
where $X_{\mu V} \equiv (X_{\mu L} + X_{\mu R})/2 $ and $X_{\mu A} \equiv  (X_{\mu R} - X_{\mu
L})/2$. Using the bound on $M_{Z^\prime}/g^\prime$ from $\bar B_s^0 - B_s^0$ mixing, we get
$a_\mu^{\mbox{\scriptsize{NP}}}  < 1.3\times 10^{-11}$ for all charge assignments, which is
smaller than the current theory uncertainty in the prediction of
$a_\mu^{\mbox{\scriptsize{SM}}}$~\cite{Agashe:2014kda}.

\subsection{Direct searches}\label{sec::directsearches}
The obvious way to search directly for a $Z^\prime$ is via 
a resonance peak in the invariant-mass distribution of its decay products. At the LHC this
experimental analysis is usually performed by the ATLAS~\cite{Aad:2012hf} 
and CMS~\cite{Chatrchyan:2012it} 
collaborations for $Z'$ production in the
$s$-channel in a rather model-independent way, but assuming validity of the narrow-width
approximation (NWA), negligible contributions of interference with the SM~\cite{Accomando:2013sfa}, and flavor-universal
$Z^\prime$ couplings to quarks. Under these assumptions, the cross section for $p p \to Z^\prime X
\to f\bar{f} X$ takes the simplified form~\cite{Carena:2004xs,Accomando:2010fz}
\begin{align}
\begin{aligned}
\sigma=\frac{\pi}{48s}&\left[c_u^f w_u\left(s,\,M_{Z^\prime}^2\right)+\,c_d^f w_d\left(s,\,M_{Z^\prime}^2\right)\right]\,,
\end{aligned}
\end{align}
where the functions $w_{u,d}$ are hadronic structure factors that encode the information of
the Drell-Yan production processes of the $Z^\prime$ and their QCD corrections (for a precise
definition of these functions we refer the reader to Ref.~\cite{Accomando:2010fz}). The
model-dependent part of the cross section is contained in the coefficients $c_{u,d}$:
\begin{align}\label{eq:cucd}
\begin{aligned}
c_u^f\simeq g^{\prime\,2}\left(X_{uL}^2+X_{uR}^2\right)\mbox{Br}\left(Z^\prime\to f\bar{f}\right)\,,\\
c_d^f\simeq g^{\prime\,2}\left(X_{uL}^2+X_{dR}^2\right)\mbox{Br}\left(Z^\prime\to f\bar{f}\right)\,.
\end{aligned}
\end{align}
While the assumptions for these expressions are not exactly fulfilled in our model, they are
applicable when neglecting the small contributions proportional to the off-diagonal CKM
matrix elements.  
The reason is that under this approximation the $Z^{\prime}$ couplings to the first two generations - which yield 
the dominant contribution to the Drell-Yan production of the $Z^{\prime}$  - are universal and flavor conserving in the quark sector, 
see Eqs.~\eqref{eq:charges_model} and~\eqref{eq:flavorstr}. Note that $X_{uL}\simeq X_{dL}$ up to
small corrections proportional to the off-diagonal CKM matrix elements that we are neglecting.\footnote{Flavor changing $Z^{\prime}$ couplings between
the first two quark generations are suppressed by $|V_{td} V_{ts}^* | \sim \lambda^5$ while violations of universality are suppressed by
$\left|V_{td}\right|^2\sim \lambda^6$ and $\left|V_{ts}\right|^2\sim \lambda^4$, with $\lambda\simeq0.226$. }

For high values of $g^\prime$ the ratio $\Gamma_{Z^\prime}/M_{Z^\prime}$ can be quite large,
spoiling the NWA. The constraint is therefore only applicable for
$g^\prime\lsim0.2$, such that $\Gamma_{Z^\prime}/M_{Z^\prime}$ does not exceed $15\%$. 
However, for the mass range accessible so far such a large coupling is typically excluded by the
constraint $R_K$, as will be shown below. Combined exclusion limits on the $Z^{\prime}$ at
$95\%$~CL using the dimuon and dielectron channels have been provided by the CMS collaboration in
the $(c_u,c_d)$ plane~\cite{Chatrchyan:2012it}. 
Using these bounds together with the constraint from $R_K$, we find that a $Z^{\prime}$ boson is excluded in our models
for $M_{Z^{\prime}} \lesssim 3$-$4$~TeV, depending on the lepton charge assignments, while $g^{\prime}$ should be $\mathcal{O}(10^{-1})$.
These bounds are stronger than those from many $Z^{\prime}$ benchmark models, for which current LHC data typically give a lower bound of
$M_{Z^\prime} \gtrsim 2$-$3$~TeV~\cite{Aad:2012hf,Chatrchyan:2012it}. The reason is a combination of a sizable value for $g'/M_{Z^\prime}$ required by
the constraint from $R_K$ and the combination of $Z^{\prime}$ couplings to fermions
appearing in our model, see Eq.~\eqref{eq:cucd}.

When the $Z^\prime$ is too heavy to be produced at a given collider, $M_{Z^\prime}^2\gg s$,
indirect searches can be carried out by looking for effects from contact interactions, \emph{i.e.}
effective operators generated by integrating out the heavy $Z^\prime$. At the LHC, the
corresponding expressions hold for $M_{Z^{\prime}} \gg 4.5$~TeV, and limits can be extracted \emph{e.g.} 
from searches for the contact interactions $(\bar q \gamma_{\mu} P_{L, R} q )  (  \bar \ell \gamma^{\mu}  P_{L,R} \ell )
$~\cite{Aad:2014wca}. These are presented in terms of benchmark scenarios considering a
single operator with a specific chirality structure. From the $95\%$~CL limit provided in
Ref.~\cite{Aad:2014wca} for the operator $(\bar q \gamma_{\mu} P_L q ) (  \bar e \gamma^{\mu}  P_R e
) $ we extract $M_{Z^\prime}/g^{\prime} \gtrsim 22$~TeV.
Searches for contact interactions of the type $(\bar e \gamma_{\mu}  P_{L, R} e )  (  \bar f
\gamma^{\mu}  P_{L,R} f )$ were performed at LEP~\cite{Schael:2013ita}. These are sensitive to a
$Z^{\prime}$-boson coupling to electrons assuming $M_{Z^{\prime}} \gg 210$~GeV and the
resulting limits are again given for benchmark scenarios. 
In this case we extract $M_{Z^{\prime}}/g^{\prime}  \gtrsim 16$~TeV from the $95\%$~CL bound on the
operator $( \bar e  \gamma^{\alpha} P_R e   )(   \bar \mu \gamma_{\alpha} P_R \mu )$. It is important to note that the extracted bounds from contact interactions using benchmark scenarios do not capture the full dynamics of our models once the $Z^{\prime}$ boson is integrated out.    Interference effects due to operators with different chirality structures will be relevant in general, a detailed analysis of these effects is however beyond the scope of this work.      These bounds
should therefore be taken as a rough guide to the sensitivity of contact interactions searches at LEP and the
LHC to the $Z^{\prime}$ of our models. However, in general they do not put relevant bounds as long as $g^{\prime}$ is not too large.

\subsection{Discriminating the different models\label{sec::models}}
The constructed class of models has two important features which imply specific deviations from
the SM:  controlled FCNCs in the down-quark sector and violations of lepton universality.
Thanks to the very specific flavor structure, the two are strongly correlated in each of our
models. While the flavor-changing couplings are universal in the models we discuss here, the
lepton charges differ. Correspondingly the models can be discriminated by processes involving
leptons, specifically rare leptonic and semileptonic decays of $B$ mesons, which can test both
features.

The effective Hamiltonian describing these $b \rightarrow s \ell^+ \ell^-$
transitions reads
\begin{align}
\mathcal{H}_{\mbox{\scriptsize eff}}=   - \frac{G_F}{\sqrt{2}}   \frac{\alpha}{\pi} V_{tb}V_{ts}^*  
\sum_{i}   \left( C_i^{\ell} \mathcal{O}^{\ell}_i  +  C_i^{\prime \ell} \mathcal{O}^{\prime \ell}_i 
\right)+\mathcal{H}_{\mbox{\scriptsize eff}}^{b\rightarrow s \gamma},
\end{align}
where %
\begin{align}
\begin{aligned}
\mathcal{O}^{\ell}_9&=\left(\overline{s}\gamma_\mu P_Lb\right)\left(\overline{\ell}\gamma^\mu
 \ell\right)\,,    &\mathcal{O}_9^{\prime \ell} &=  \left(\overline{s}\gamma_\mu P_Rb\right)\left(\overline{\ell}\gamma^\mu \ell\right)\,,\\[5pt] \mathcal{O}^{\ell}_{10}&= \left(\overline{s}\gamma_\mu P_Lb\right)\left(\overline{\ell}\gamma^\mu\gamma_5 \ell\right)\,,   &\mathcal{O}_{10}^{\prime \ell}&= \left(\overline{s}\gamma_\mu P_Rb\right)\left(\overline{\ell}\gamma^\mu\gamma_5 \ell\right)\,,    \\[5pt]
\mathcal{O}^{\ell}_{S}&= m_b (  \bar s P_R b )( \bar \ell \ell) \,,  &\mathcal{O}^{\prime
 \ell}_{S}&= m_b (  \bar s P_L b )( \bar \ell \ell) \,,  \\[5pt] \mathcal{O}^{\ell}_{P}&= m_b (  \bar s P_R b )( \bar \ell  \gamma_5 \ell) \,,  &\mathcal{O}^{\prime \ell}_{P}&= m_b (  \bar s P_L b )( \bar \ell  \gamma_5 \ell) \,.
\end{aligned}
\end{align}
The piece $\mathcal{H}_{\mbox{\scriptsize eff}}^{b\rightarrow s \gamma}$ is the effective
Hamiltonian for the radiative transitions including dipole operators~\cite{Buras:2012jb}. New
physics contributions to $\mathcal{H}_{\mbox{\scriptsize eff}}^{b\rightarrow s \gamma}$ arise in our
models from one-loop diagrams mediated by $Z^{\prime}$ and Higgs bosons. Both of them are
very small due to the assumed mass scale of several TeV~\cite{Buras:2012jb}.

At the $b$-quark scale, the SM contribution to the remaining part reads
$C_{9}^{\mbox{\scriptsize{SM}}} \simeq - C_{10}^{\mbox{\scriptsize{SM}}} \simeq  4.2\,\forall\,\ell$, while all other contributions are
negligible. The chirality-flipped operators $\mathcal{O}^{\prime}_{9,10}$ receive negligible
contributions also in our models since $\epsilon^{\prime\,d}_{R,sb}=0$, while all others receive
contributions from $Z^\prime$ or Higgs exchanges ~\cite{Buras:2012jb,Buras:2013uqa}.
The $Z^{\prime}$ contribution to $C_{9,10}^{\ell}$ is given by
\begin{align}
\begin{aligned}
C_{9}^{\mbox{\scriptsize{NP}}  \ell}\simeq&     - \frac{\pi v^2   }{\alpha V_{ts}^* V_{tb} }  
\frac{ \epsilon^{\prime\,d}_{L,sb}  \epsilon^{\prime\,e}_{V,\ell \ell}   }{M_{Z^{\prime}}^2}     \,, \\  C_{10}^{\mbox{\scriptsize{NP}}  \ell}\simeq&     - \frac{\pi v^2   }{\alpha V_{ts}^* V_{tb} }   \frac{ \epsilon^{\prime\,d}_{L,sb}  \epsilon^{\prime\,e}_{A,\ell \ell}   }{M_{Z^{\prime}}^2}               \,,
\end{aligned}
\end{align}
where $C_{i}^{\ell} \equiv C_{i}^{\mbox{\scriptsize{SM}}}  +
C_{i}^{\mbox{\scriptsize{NP}} \ell}$.
In Table~\ref{tab:correlations} we show the correlations between the Wilson coefficients
$C_{9,10}^{\mbox{\scriptsize{NP}}  \ell}$ in our models and provide 
$C_{9}^{\mbox{\scriptsize{NP}}  \mu}$ as a function of $g^{\prime}/M_{Z^{\prime}}$ by
introducing a conveniently normalized parameter $\kappa_{9}^{\mu}$:
\be
 C_9^{\mbox{\scriptsize{NP}} \mu}  \equiv  \kappa_9^{\mu}   \times 10^{4} \left(   \frac{ g^{\prime}
 v }{  M_{Z^{\prime}}  }  \right)^2=\kappa_9^{\mu}   \times605~{\rm TeV}^2 \left(   \frac{ g^{\prime}
 }{  M_{Z^{\prime}}  }  \right)^2 \,.
\ee
This allows the direct estimation of $C_{9,10}^{\mbox{\scriptsize{NP}}  \ell}$ in terms of
$g^{\prime}/M_{Z^{\prime}}$, useful for phenomenological purposes.

%%%%%%%%%%%%%%%%%%%%%%%%%%%%%%%%%%%%%%%%%%%%%%%%%%%%%%%%%%%%%%%%%%%%%%%%%%
\begin{table}
\begin{center}
\caption{\it \small Correlations among the NP contributions to the effective operators
$\mathcal{O}_{9,10}^{\ell}$.} \vspace{0.2cm}  \tabcolsep 0.03in
\begin{tabular}{|c|c|c|c|c|} \rowcolor{RGray}
\hline 
Model   & $C_{10}^{\mbox{\scriptsize{NP}} \mu}/C_{9}^{\mbox{\scriptsize{NP}}\mu}$ &  $C_{9}^{\mbox{\scriptsize{NP}}e}/C_{9}^{\mbox{\scriptsize{NP}}\mu}$  & $C_{10}^{\mbox{\scriptsize{NP}}e}/C_9^{\mbox{\scriptsize{NP}}\mu}$   &    $\kappa_{9}^{\mu}$    \\[0.1cm] 
\hline
(1,2,3) & $3/17$  & $9/17$                 & $3/17$   & $-1.235$ \\[0.1cm] \rowcolor{CGray}  
(1,3,2) & $0$  & $-9/8$ & $-3/8$    &  $0.581$  \\[0.1cm] 
(2,1,3) & $1/3$  & $17/9$                 & $1/3$   & $-0.654$ \\[0.1cm] \rowcolor{CGray}
(2,3,1) & $0$ & $-17/8$ & $-3/8$    &  $0.581$  \\[0.1cm]   
(3,1,2) & $1/3$  & $-8/9$                 & $0$  & $-0.654$ \\[0.1cm] \rowcolor{CGray}
(3,2,1) & $3/17$ & $-8/17$                 & $0$      & $-1.235$ \\[0.1cm] 
\hline
\end{tabular}
\label{tab:correlations}
\end{center}
\end{table}
%%%%%%%%%%%%%%%%%%%%%%%%%%%%%%%%%%%%%%%%%%%%%%%%%%%%%%%%%%%%%%%%%%%%%%%%%%

%
The CP-even Higgs $H_2$ and the CP-odd Higgs $A$ can give sizable contributions to
$O_{S}^{\ell}$ and $O_{P}^{\ell}$, respectively. For muons we have\footnote{The correlation $C_S = -C_P$ 
was expected given the assumed decoupling in the scalar sector~\cite{Alonso:2014csa}.}
\be
C_S^{\mbox{\scriptsize{NP}}  \mu} \simeq - C_P^{ \mbox{\scriptsize{NP}} \mu}  \simeq \dfrac{ -  2 
\, \pi  \,  ( t_{\beta}  +  t_{\beta}^{-1} )  (N_{\ell})_{\mu \mu}   }{   \alpha  \bar M_{H}^2    } \,.
\ee
Here we have denoted the quasi-degenerate masses of $H_2$ and $A$ by $\bar M_{H} \equiv M_{H_2}
\simeq M_A$, see Appendix~\ref{app:scalar} for details of the scalar sector.  The coupling
$(N_\ell)_{\mu\mu}$ is again model-dependent:
$(N_{\ell})_{\mu \mu} =t_{\beta} m_{\mu}$ for models $(1,2,3)$, $(2,1,3)$, $(3,1,2)$ and $(3,2,1)$,
while $(N_{\ell})_{\mu \mu} = - t^{-1}_{\beta} m_{\mu}$ for $(1,3,2)$ and $(2,3,1)$.  The
suppression by the muon mass renders these contributions negligible, apart from
observables in which also the SM and $Z^\prime$ contributions receive this suppression, \emph{e.g.}
$B_s\to\mu^+\mu^-$.
Higgs contributions to $O_{S, P}^{\prime \ell}$ will be additionally suppressed by a factor
$m_s/m_b$ compared to the corresponding non-primed operators and are neglected in the following.

The angular distributions of neutral-current semileptonic $b \rightarrow s \ell^+ \ell^-$
transitions allow for testing these coefficients in global fits. Furthermore, they provide
precise tests of lepton universality when considering ratios of the type 
\be  \label{cpo}
R_M \equiv   \frac{\mathrm{Br}(   \bar B \rightarrow \bar M \mu^+ \mu^- )}{\mathrm{Br}(   \bar B
\rightarrow \bar M e^+ e^- )}\stackrel{\rm SM}{=}1+\mathcal O(m_\mu^2/m_b^2)\,,
\ee
with $M\in\{K, K^*, X_s, K_0(1430),\ldots\}$~\cite{Hiller:2003js},
where many sources of uncertainties cancel when integrating over identical phase-space regions.

Additional sensitivity to the dynamics of NP can be obtained via
double-ratios~\cite{Hiller:2014ula}, \be \label{ew:dr}
\widehat R_{M} \equiv \frac{ R_M}{R_K} \,.%
\ee%
It was shown in Ref.~\cite{Hiller:2014ula} that the dependence on the NP coupling to left-handed
quarks cancels out in $\widehat R_{M}$.  Since $C_{9,10}^{\prime \ell} =0$ in our models we have
$\widehat R_{K^*} = \widehat  R_{X_s} = \widehat  R_{K_0(1430)} =1$, providing an important test of
the flavor structure of our models. 

The possible hadronic final states for these ratios yield sensitivity to different NP
structures, thereby providing complementary information.
The discriminating power of these observables has been shown recently in
Ref.~\cite{Varzielas:2015iva} within the framework of leptoquark models and in more general
contexts in Refs.~\cite{Hiller:2014yaa,Hiller:2014ula,Blake:2015tda}.

To analyze the deviations from flavor-universality we define $R_M\simeq  1 + \Delta  + \Sigma$,
where 
$\Sigma$ is the pure contribution from NP and $\Delta$ the one from the interference
with the SM. These quantities are given by
\begin{align}
\begin{aligned}
\Delta =&\, \frac{2}{    |C_{9}^{\mbox{\scriptsize{SM}}}|^2 + | C_{10}^{\mbox{\scriptsize{SM}}} |^2 }  \left[   \mathrm{Re}\left\{       C_{9}^{\mbox{\scriptsize{SM}}}    \left({ C_{9}^{\mbox{\scriptsize{NP}} \mu}}\right)^*  \right\}  \right.\\
 &\left.+ \,  \mathrm{Re}\left\{   C_{10}^{\mbox{\scriptsize{SM}}}    \left({ C_{10}^{\mbox{\scriptsize{NP}} \mu}}\right)^*  \right\}   - \left(\mu \rightarrow e\right) \right] \,, \\[0.1cm]
 \Sigma =&\, \frac{  \left|C_{9}^{\mbox{\scriptsize{NP}}\mu}\right|^2 + \left|
 C_{10}^{\mbox{\scriptsize{NP}}\mu} \right|^2}{  \left|C_{9}^{\mbox{\scriptsize{SM}}}\right|^2 + \left| C_{10}^{\mbox{\scriptsize{SM}}} \right|^2} -   (\mu \rightarrow e) \,,
\end{aligned}
\end{align}
and are valid to a very good approximation given the present experimental uncertainties.

Our models also give clean predictions for the rare decay modes $B \rightarrow \{K, K^*, X_s\}
\nu \bar \nu$~\cite{Buras:2012jb}.  Tree-level $Z^{\prime}$-boson contributions to these decays
are generically expected of the same size as in $b\to s\ell\ell$ transitions. However, the
modes with neutrino final states do not distinguish between our different models, since there
is no sensitivity to the neutrino species.  We obtain again a universal value for all ratios
$R_M^\nu=\mathrm{Br}(B\to M\bar\nu\nu)/\mathrm{Br}(B\to M\bar\nu\nu)|_{\rm SM}$,
due to the fact that $\epsilon^{\prime\,d}_{R,sb}=0$. For the same reason
the average of the $K^*$ longitudinal polarization fraction in $B \rightarrow K^* \nu \bar \nu$ is
not affected by the $Z^{\prime}$ exchange contribution~\cite{Buras:2012jb}.
The enhancement of $R_M^\nu$ turns out to be relatively small, $\mathcal{O}(10\%)$ 
for $g^{\prime} \sim 0.1$ and $M_{Z^{\prime}} \sim \mathcal{O}(\text{TeV})$. The reason is again the sum over
the different neutrino species: due to the different charges, the $Z^\prime$ contribution interferes
both constructively and destructively, leaving a net effect that is smaller than in the modes with
charged leptons.

The leptonic decays $B_{s,d}^{0} \rightarrow \ell^+ \ell^-$ constitute another sensitive
probe of small NP effects. Within the SM, these decays arise again at the loop level
and are helicity suppressed. Due to the leptonic final state, the theoretical prediction of these
processes is very clean. 
They receive $Z^{\prime}$- and Higgs-mediated contributions at tree-level in our model through
the operators $\mathcal{O}_{10, S, P}^{\mu}$~\cite{Hiller:2014yaa},
\begin{align}
\begin{split}
\dfrac{\mathrm{Br}(  \bar B_s \rightarrow \mu^+ \mu^- )}{ {\mathrm{Br}(\bar B_s \rightarrow \mu^+
\mu^- )^{\mbox{\scriptsize{SM}}} }       } \simeq&   \left|  1- 0.24 C_{10}^{\mbox{\scriptsize{NP}}
\mu}    - y_{\mu}    C_{P}^{\mbox{\scriptsize{NP}}  \mu}    \right|^2  \,   \\   &+ | y_{\mu}  
C_{S}^{\mbox{\scriptsize{NP}}  \mu}  |^2  \,,
\end{split}
\end{align}
where $y_{\mu } \simeq 7.7  \,m_b$. For the range of model parameters relevant here, the
$Z^{\prime}$ contribution to  $\bar B_s \rightarrow \mu^+ \mu^-$ is very small with respect to
the SM, $\sim 1\%$ for $M_{Z^{\prime}} \sim 5$~TeV and $g^{\prime} \sim 0.1$. 
Larger contributions are possible
due to scalar mediation 
for the models $(1,2,3)$, $(2,1,3)$, $(3,1,2)$ and $(3,2,1)$, given that $C_S \simeq - C_P
\propto t_{\beta}^2$ in these cases. Taking $\bar M_{H} \sim 10$~TeV and $t_{\beta} \sim 30$
for example, one obtains a suppression of $\mathrm{Br}(  \bar B_s \rightarrow \mu^+ \mu^- )$ of about $10\%$ relative
to the SM. 
In our model the ratio $\mathrm{Br}(  \bar B_d
\rightarrow \mu^+ \mu^- )/\mathrm{Br}(  \bar B_s \rightarrow \mu^+ \mu^- )$ remains unchanged with
respect to the SM to a very good approximation.

If a $Z^{\prime}$ boson is discovered during the next runs of the LHC~\cite{Godfrey:2013eta},
its decays to leptons can be used to discriminate the models presented here.    We define the
ratios 
\begin{align}
\mu_{f/f^{\prime}}&\equiv \frac{\sigma(pp\rightarrow Z^\prime\rightarrow
f\bar{f})}{\sigma(pp\rightarrow Z^\prime\rightarrow  f^{\prime} \bar  f^{\prime})}\,,
\end{align}
where again the dominant sources of uncertainty cancel, rendering them 
particularly useful to test lepton universality.  This possibility 
has also been discussed in
Refs.~\cite{Demir:2005ti}.
In our models we have
\begin{align}
\mu_{b/t} \simeq  \frac{X_{bL}^2+X_{bR}^2}{X_{tL}^2+X_{tR}^2} \,, \qquad \mu_{\ell/\ell^{\prime}}  \simeq   \frac{X_{\ell L}^2+X_{\ell R}^2}{X_{\ell^{\prime} L}^2+X_{\ell^{\prime} R}^2} \,.
\end{align}
The first ratio is fixed in our models, $\mu_{b/t} \simeq 1$, while large deviations from
$\mu_{\ell/\ell^{\prime}}=1$ are possible. The model-specific predictions for the
ratio $\mu_{\ell/\ell^{\prime}}$ are shown in Table~\ref{tab:correlationsLHC}. We note that
there is a correlation between $R_K$ and $\mu_{e/\mu}$:  an enhancement of $R_K$ (as
present in models $(1,3,2)$, $(2,1,3)$ and $(2,3,1)$) implies an enhancement of $\mu_{e/\mu}$ and
the other way around.
Violations of lepton universality 
can therefore be tested by complementary measurements of flavor observables at LHCb or Belle~II
and $Z^{\prime}$ properties to be measured at ATLAS and CMS.

%%%%%%%%%%%%%%%%%%%%%%%%%%%%%%%%%%%%%%%%%%%%%%%%%%%%%%%%%%%%%%%%%%%%%%%%%%
\begin{table}
\begin{center}
\caption{\it \small   Model-dependent predictions for the ratio $\mu_{\ell/\ell^{\prime}}$.}
\vspace{0.2cm}  \tabcolsep 0.03in
\begin{tabular}{|c|c|c|c|} \rowcolor{RGray}
\hline 
Model   & $\mu_{e/\mu}$    & $\mu_{e/\tau}$    & $\mu_{\mu/\tau}$     \\[0.1cm] 
\hline
(1,2,3) & $45/149$  &     $45/32$  &   $149/32$  \\[0.1cm] \rowcolor{CGray}  
(1,3,2) & $45/32$  &   $45/149$ &   $32/149$ \\[0.1cm] 
(2,1,3) & $149/45$  &   $149/32$ &   $45/32$  \\[0.1cm] \rowcolor{CGray}
(2,3,1) & $149/32$  &   $149/45$  &  $32/45$   \\[0.1cm]   
(3,1,2) & $32/45$  &   $32/149$ &   $45/149$ \\[0.1cm] \rowcolor{CGray}
(3,2,1) & $32/149$  &     $32/45$ &    $149/45$   \\[0.1cm] 
\hline
\end{tabular}
\label{tab:correlationsLHC}
\end{center}
\end{table}
%%%%%%%%%%%%%%%%%%%%%%%%%%%%%%%%%%%%%%%%%%%%%%%%%%%%%%%%%%%%%%%%%%%%%%%%%%

\subsection{Interpretation of \texorpdfstring{$b \rightarrow s \ell^+ \ell^-$}{the}
anomalies  \label{npsec} }
The LHCb collaboration recently performed two measurements of decays with $b\to s\ell^+\ell^-$
transitions that show a tension with respect to the SM expectations. The first measurement is that
of the ratio $R_K$ introduced in the last section, where LHCb measures~\cite{Aaij:2014ora}
$R_K^{\mbox{\scriptsize{LHCb}}} = 0.745^{+0.090}_{-0.074}\pm0.036$ for $q^2\in[1,6]~{\rm GeV}^2$,
a $2.6\,\sigma$ deviation with respect to the SM prediction~\cite{Hiller:2003js,
Bobeth:2007dw,Bouchard:2013mia}.
The second measurement is the angular analysis of $B\to K^*\mu^+\mu^-$ decays~\cite{Aaij:2013qta},
where an excess of $2.9\sigma$ is observed for the angular observable $P_5^{\prime}$ in the bin
$q^2\in[4,6]~{\rm GeV}^2$,\footnote{The bin $q^2\in[6,8]~{\rm GeV}^2$ shows nominally a tension
with identical significance; however, this bin is considered less theoretically clean, due to the
larger influence of charm resonances.} which has been constructed to reduce the influence of form
factor uncertainties~\cite{Descotes-Genon:2013vna}.

\begin{figure}[ht!]
\centering
\includegraphics[width=8cm,height=8cm]{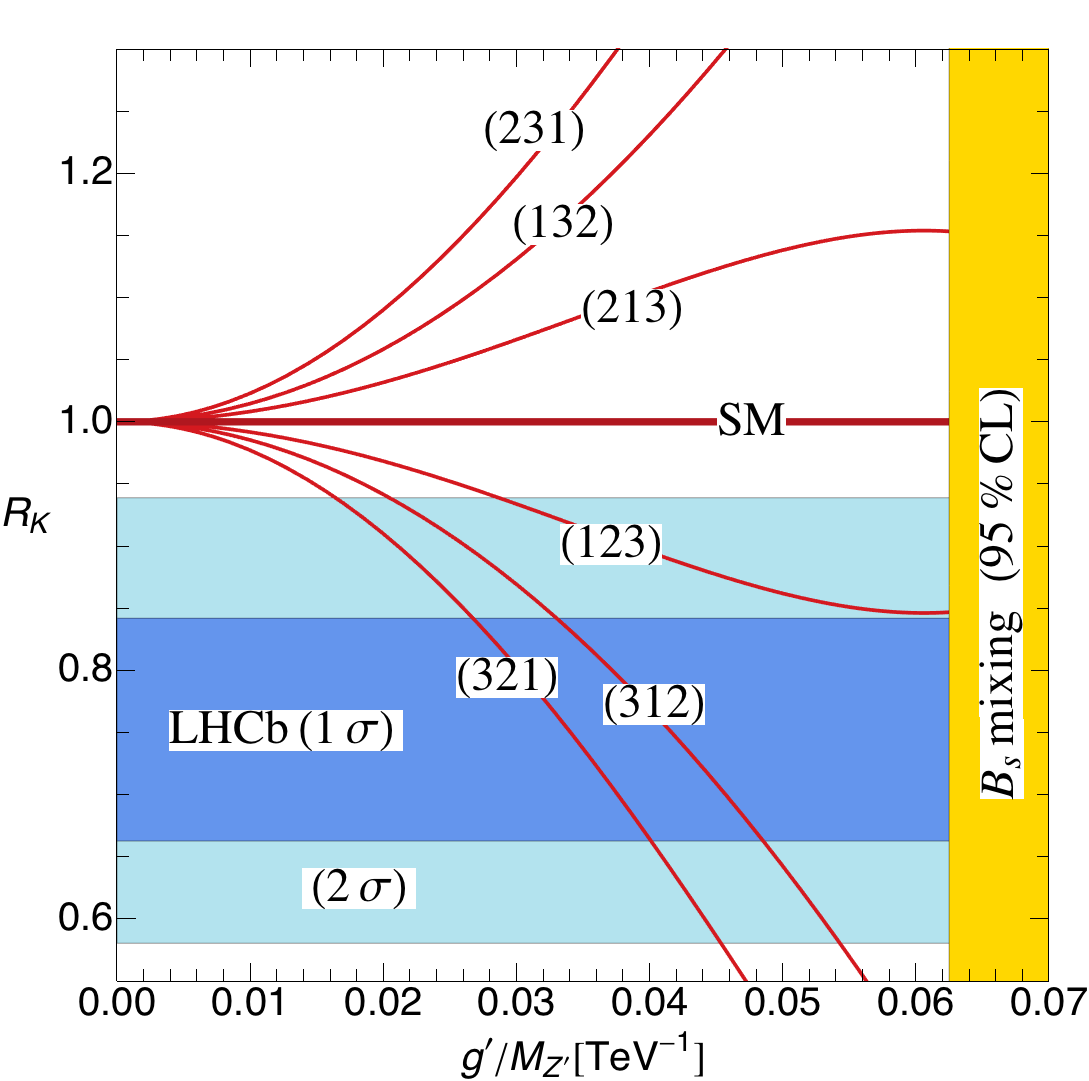}
\caption{ \it \small  Model-dependent predictions for $R_{K}$ as a function of
$g^{\prime}/M_{Z^{\prime}}$. The recent measurement of $R_K$ by the LHCb collaboration is shown at
$1 \sigma$ and $2 \,\sigma$.   Constraints from $B_s$ mixing are also shown at $95\%$~CL. }\label{fig:RKmodels}
\end{figure}

Attributing the measurement of $R_K$ solely to NP, \emph{i.e.} assuming that it will be confirmed
with higher significance in the next run of the LHC and/or by additional $R_M$ measurements, it
is the first way to exclude some of our models, since it requires sizable non-universal contributions
with a specific sign. The flavor structure in each of our models is fixed, so half of them cannot
accommodate $R_K< 1$, namely $(1,\,3,\,2)$, $(2,1,3)$ and $(2,\,3,\,1)$. This is illustrated in
Fig.~\ref{fig:RKmodels}, where it is additionally seen that a large deviation from $R_K=1$, as
indicated by the present central value and $1\sigma$ interval, can actually only be explained in two
of the remaining models. This strong impact shows the importance of further measurements of $R_M$
ratios.

\begin{figure*}
\includegraphics[width=5.7cm,height=5.7cm]{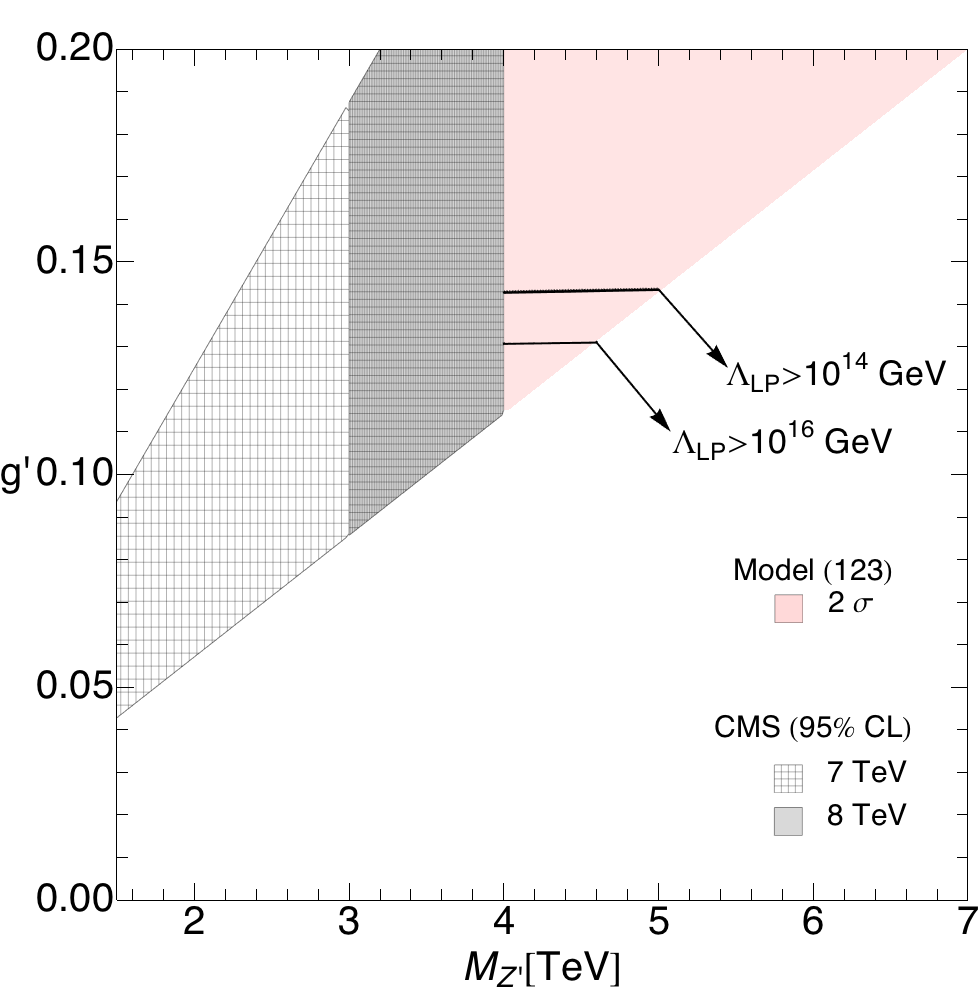}
~
\includegraphics[width=5.7cm,height=5.7cm]{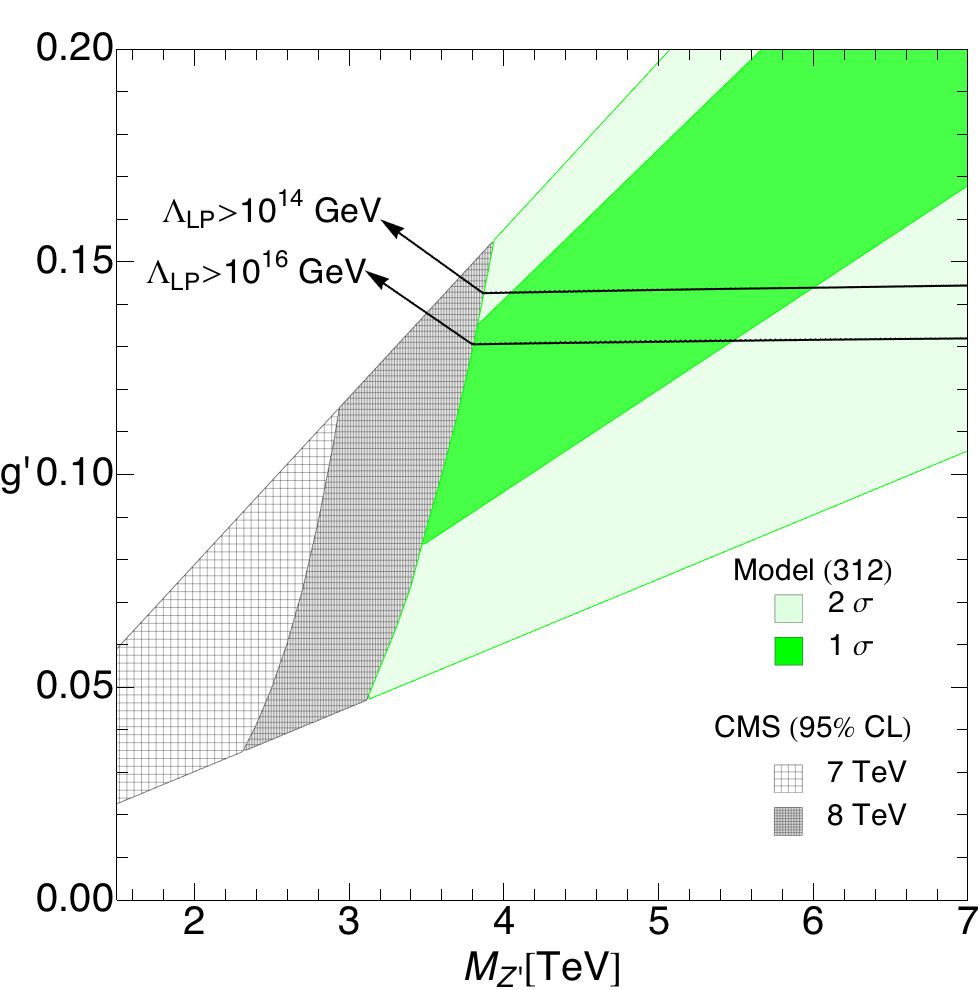}
~
\includegraphics[width=5.7cm,height=5.7cm]{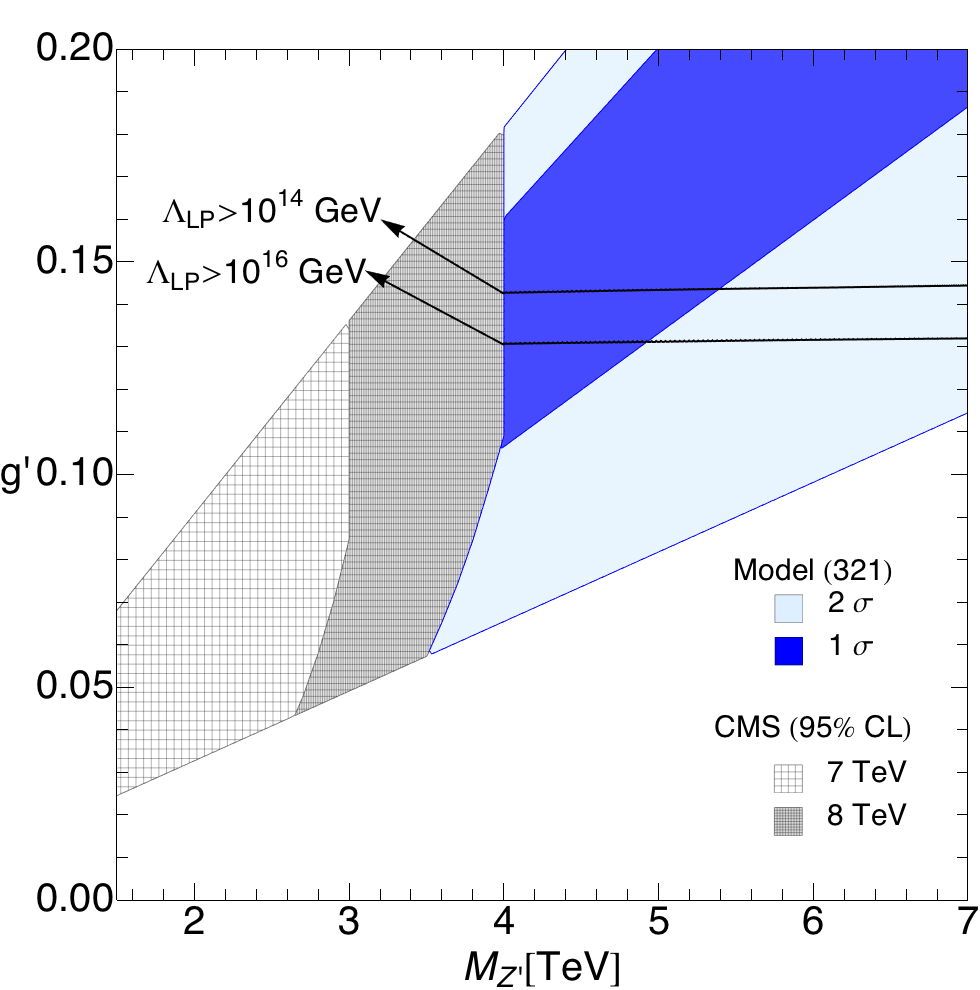}
\caption{\it \small  Regions allowed at $1\sigma$ and $2\sigma$ by the $R_K$ measurement in the
$\{M_{Z^{\prime}},g^{\prime}\}$ plane for the models $(1,2,3)$, $(3,1,2)$ and $(3,2,1)$.   Exclusion
limits from $Z^{\prime}$ searches at the LHC are shown in gray. The black lines indicate bounds from
perturbativity of $g^\prime$.}  \label{fig:constraints}
\end{figure*}

In Fig.~\ref{fig:constraints} we show the constraints from the $R_K$
measurement for the remaining models $(1,2,3)$, $(3,1,2)$ and $(3,2,1)$. The allowed
regions are consistent with the constraint from $B_s^0$-meson mixing. LHC searches
for a $Z^{\prime}$ boson exclude values of $M_{Z^{\prime}}$ below $3$-$4$~TeV, as discussed in
Sec.~\ref{sec::directsearches}; the corresponding areas are shown in gray. We also show the
theoretical perturbativity bounds obtained from the requirement that the Landau pole for the $\mathrm{U(1)}^\prime$ gauge coupling appears beyond
the see-saw or the Grand Unification scales, \emph{i.e.}
$\Lambda_{\mbox{\scriptsize{LP}}}>10^{14}$~GeV and $\Lambda_{\mbox{\scriptsize{LP}}}>10^{16}$~GeV, respectively.

Regarding the angular analysis in $B\to K^*\mu^+\mu^-$, the situation is more complicated. The
wealth of information provided in this and related measurements requires a global analysis, which is
beyond the scope of this work. However, a number of model-independent studies have been carried out
\cite{Descotes-Genon:2013wba,Beaujean:2013soa,Hurth:2013ssa,Ghosh:2014awa,
Hurth:2014vma,Altmannshofer:2015sma}. 
These analyses differ in terms of
the statistical methods used, treatment of hadronic uncertainties (see also
Ref.~\cite{Jager:2014rwa}), and consequently in the significance they find for the NP
hypothesis over the SM one. However, they agree that a contribution $C_9^{{\rm NP}\mu}\sim -1$ can
fit the data for $P_5^\prime$ without conflicting with other observables. Additional contributions
from \emph{e.g.} $C_{10}^{{\rm NP}\mu}$ can be present, but are less significant and tend to be
smaller. When using Table~\ref{tab:correlations} to translate the $R_K$ measurement into a bound
on $C_9^{{\rm NP}\mu}$ in our models, see Table~\ref{tab:correlationsII}, the ranges are perfectly
compatible with the values obtained from $P_5^\prime$, as also observed for other $Z^\prime$
models~\cite{Descotes-Genon:2013wba,Hurth:2013ssa,Altmannshofer:2015sma}.
This is highly non-trivial given the strong correlations in our models and will allow for decisive
tests with additional data.

%%%%%%%%%%%%%%%%%%%%%%%%%%%%%%%%%%%%%%%%%%%%%%%%%%%%%%%%%%%%%%%%%%%%%%%%%%
\begin{table}
\begin{center}
\caption{\it \small  Model-dependent bound on $C_{9}^{\mbox{\rm{\scriptsize{NP}}}\mu}$
from the $R_K$ measurement.   The constraint from $B_s$ mixing is taken into account. } \vspace{0.2cm}  \tabcolsep 0.03in
\begin{tabular}{|c|c|c|c|c|c|} \rowcolor{RGray}
\hline 
Model   & $C_{9}^{\mbox{\scriptsize{NP}}\mu} (1\sigma)$  & $C_{9}^{\mbox{\scriptsize{NP}}\mu}  (2 \sigma)$     \\  \hline 
(1,2,3)   &  -- & $[-2.92, -0.61]$    \\[0,1cm]
(3,1,2) & $[-0.93, -0.43]$  &  $[-1.16, -0.17]$  \\[0.1cm] 
(3,2,1) & $[-1.20, -0.53]$  & $[-1.54, -0.20]$ \\
\hline
\end{tabular}
\label{tab:correlationsII}
\end{center}
\end{table}
%%%%%%%%%%%%%%%%%%%%%%%%%%%%%%%%%%%%%%%%%%%%%%%%%%%%%%%%%%%%%%%%%%%%%%%%%%

Let us further comment on the implications of the recent measurements of
$\bar B_{d,s}^0 \rightarrow \mu^+ \mu^-$ decays~\cite{CMS:2014xfa},
\begin{align}
\begin{aligned}
\dfrac{\mathrm{Br}(  \bar B_s \rightarrow \mu^+ \mu^- )^{\mbox{\scriptsize{exp}}}}{
{\mathrm{Br}(\bar B_s \rightarrow \mu^+ \mu^- )^{\mbox{\scriptsize{SM}}} }       }
&=0.76^{+0.20}_{-0.18}\,,\\
\dfrac{\mathrm{Br}(  \bar B_d \rightarrow \mu^+ \mu^- )^{\mbox{\scriptsize{exp}}}}{
{\mathrm{Br}(\bar B_d \rightarrow \mu^+ \mu^- )^{\mbox{\scriptsize{SM}}} }       }
&=3.7^{+1.6}_{-1.4}\,.
\end{aligned}
\end{align}
Both results seem to hint at a deviation from the SM, however in opposite directions. A
confirmation of this situation with higher significance would rule out our models which predict
these ratios to be equal; however, the uncertainties in the $B_d$ mode are still large. Regarding
the $B_s$ mode, which is measured consistent with the SM prediction, it should be noted that the
result depends sensitively on the value adopted for $|V_{cb}|$, where the value from inclusive $B\to
X_c\ell\nu$ decays was chosen in the SM calculation~\cite{Bobeth:2013uxa}. In any case, as discussed
in Sec.~\ref{sec::models}, a potential shift could be explained in the context of our models by
scalar contributions, which are, however, not directly related to the $Z^\prime$ contributions
discussed above.

The models that accommodate the $b \rightarrow s \ell^+ \ell^-$
data can be further discriminated using the observables described in the last section,
notably using the ratios $\mu_{\ell/\ell^{\prime}}$, provided a $Z^{\prime}$ boson is discovered
at the LHC.

\section{Conclusions}\label{sec:con}

The class of family-non-universal $Z^\prime$ models presented in this article exhibits FCNCs at tree
level that are in accordance with available flavor constraints while still inducing potentially
sizable effects in various processes, testable at existing and future colliders. This is achieved by
gauging the specific (BGL-)symmetry structure, introduced in Ref.~\cite{Branco:1996bq} for the first
time, which renders the resulting models highly predictive.

The particle content of the models is minimal in the sense that the extension of the $\mathrm{U(1)}^\prime$
symmetry to the lepton sector allows to restrict the fermion content to the SM one. The only
additional particles are then a second Higgs doublet, a scalar singlet and the $Z^\prime$ boson,
all of which are heavy after spontaneous symmetry breaking, due to the large mass scale for the
singlet. The anomaly conditions largely determine the charge assignments under the $\mathrm{U(1)}^\prime$
symmetry; the remaining freedom is used for two phenomenologically motivated choices, leaving only
six possible models which are related by permutations in the lepton sector.

The main phenomenological features of these models can be summarized as follows:
\begin{enumerate}
  \item FCNCs at tree level in the down-quark sector, which are mediated by heavy $Z^\prime$ gauge
  bosons and neutral scalars, controlled by combinations of CKM matrix elements and/or
  fermion mass factors.
  \item Non-universal lepton couplings determined by the charges under the additional
  $\mathrm{U(1)}^\prime$ symmetry.
  \item No FCNCs in the charged-lepton or up-quark sectors.
  \item Complete determination of the $\mathrm{U(1)}^\prime$ sector up to two real parameters,
  $M_{Z^\prime}$ and $g^\prime$, where all observables at the electroweak scale and below depend
  only on the combination $g^\prime/M_{Z^\prime}$.
\end{enumerate}

Present data already strongly restrict the possible parameter ranges in our models: direct searches
exclude $Z^\prime$ masses below $3-4$~TeV and the constraint from $B$ mixing implies
$M_{Z^\prime}/g^\prime\geq 16$~TeV ($95\%$ CL). 
Theoretical bounds from perturbativity give an upper bound on the value of the gauge coupling,
\emph{e.g.} $g^\prime\lsim0.14$ for a Landau pole beyond the see-saw scale.
Nevertheless, three of our models can explain the deviations from SM expectations in $b\to
s\ell^+\ell^-$ transitions seen in LHCb measurements \cite{Aaij:2013qta,Aaij:2014ora}, while the
other three are excluded (at $95\%$~CL) by $R_K<1$. 
These findings are illustrated in Figs.~\ref{fig:RKmodels} and~\ref{fig:constraints}. 

Any significant deviation from the SM in an observable with $Z^\prime$ contributions allows
for predicting all other observables, for instance the values for
$C_{9,10}^{\mbox{\scriptsize{NP}}\ell}$ obtained from $R_K$, see Tables~\ref{tab:correlations}
and~\ref{tab:correlationsII}. Further characteristic predictions in our models include the
following:

\begin{itemize}

\item The absence of flavor-changing $Z^{\prime}$ couplings to right-handed quarks implies
$ \widehat  R_M =1$, \emph{i.e.} all ratios defined in Eq.~\eqref{cpo} are
expected to be equal, $R_K=R_{K^*}=R_{X_s}=\ldots$
The same holds for the ratios $R_M^\nu$ in $B\to M\bar \nu\nu$ decays.

\item If a $Z^\prime$ is discovered at the LHC,
measurements of  $\sigma(pp \rightarrow Z^{\prime} \rightarrow \ell_i \bar \ell_i )/\sigma(pp
\rightarrow Z^{\prime} \rightarrow \ell_j\bar \ell_j )$ can be used to discriminate between our
models, due to the specific patterns of lepton-non-universality, see
Table~\ref{tab:correlationsLHC}.

\item Leptonic down-quark FCNC decays are sensitive to the Higgs couplings in our model,
specifically $B_{d,s}\to \ell^+\ell^-$. Double ratios of $B_s$ and $B_d$ decays are again expected
to equal unity.

\end{itemize}

In the near future we will therefore be able to differentiate our new class of models from other
$Z^\prime$ models as well as its different realizations from each other. This will be possible due
to a combination of direct searches/measurements at the LHC and high-precision measurements at low
energies, \emph{e.g.} from Belle~II and LHCb. Further progress can come directly from theory,
\emph{e.g.} by more precise predictions for $\Delta m_{d,s}$ or $\epsilon_K$.

As a final remark, we recall that in this minimal implementation of gauged BGL symmetry neutrinos
are massless. Additional mechanisms to explain neutrinos masses can introduce new solutions to the
anomaly equations; these new variants are subject to future work.

\begin{acknowledgments}
The work of A.C. is supported by the Alexander von Humboldt Foundation.  
The work of J.F. has been supported in part by the Spanish Government and ERDF from the EU
Commission [Grants No. FPA2011-23778, FPA2014-53631-C2-1-P, No.~CSD2007-00042 (Consolider Project
CPAN)]. J.F.  also acknowledges VLC-CAMPUS for an ``Atracci\'{o} del Talent"  scholarship.  
The work of M.J. was financially supported by the ERC Advanced Grant project ``FLAVOUR'' (267104)
and the DFG cluster of excellence ``Origin and Structure of the Universe''.
H.S.'s work was supported by the National Research Foundation of Korea (NRF) grant funded by the
Korea government (MEST) (No. 2012R1A2A2A01045722).
 A.C. is grateful to A.~Vicente and J.~Virto for discussions about the $b \rightarrow s \ell^+
\ell^-$ anomalies.
\end{acknowledgments}

\begin{appendix}

\section{Anomaly cancellation conditions} \label{sec:anomaly}
In this appendix we present the restrictions on the charges derived from the requisite gauge
anomaly cancellation for quarks transforming according to Eq.~\eqref{eq:system}
and allowing the leptons to transform in the most general way:
\begin{align}
\begin{split}
\mathcal{X}_L^\ell&=\mathrm{diag}\left(X_{eL},X_{\mu L},X_{\tau L}\right)\,,\\
\mathcal{X}_R^e&=\mathrm{diag}\left(X_{eR},X_{\mu R},X_{\tau R}\right)\,.
\end{split}
\end{align}

As mentioned in Sec.~\ref{s:frame}, the QCD anomaly condition
is automatically satisfied within BGL models. On the other hand, the anomaly
conditions involving a single $\mathrm{U(1)}^\prime$ gauge boson together with
$\mathrm{SU(2)}_L$ or $\mathrm{U(1)}_Y$ ones, and the mixed
gravitational-$\mathrm{U(1)}^\prime$ anomaly are non-trivial and read 
\begin{align}
\begin{aligned}
\mathcal{A}_{221^\prime}=&\, 3 X_{uR}+\frac{3}{2} X_{tR}+\frac{9}{2} X_{dR}+\sum_{\alpha=e,\mu,\tau}X_{\alpha L}\,,\\
\mathcal{A}_{111^\prime}=&\,-\frac{5}{2} X_{uR}-\frac{5}{4} X_{tR}-\frac{3}{4} X_{dR}\\
&+\sum_{\alpha=e,\mu,\tau}\left(\frac{1}{2}X_{\alpha L}-X_{\alpha R}\right)\,,\\
\mathcal{A}_{GG1^\prime}=& \sum_{\alpha=e,\mu,\tau}\left(2X_{\alpha L}-X_{\alpha R}\right)\,.
\end{aligned}
\end{align}
The triangle diagrams involving two or three $\mathrm{U(1)}^\prime$ gauge bosons result in
more complicated conditions for the BGL charges, which include a quadratic and a cubic
equation:
\begin{align}
\mathcal{A}_{11^\prime 1^\prime}=&\,-\frac{7}{2}X_{uR}^{2}-\frac{7}{4}X_{tR}^2+\frac{15}{4}
X_{dR}^{2}+X_{uR}X_{dR}\nonumber\\
&+\frac{1}{2}X_{tR}X_{dR}-\sum_{\alpha=e,\mu,\tau}\left(X_{\alpha L}^2-X^2_{\alpha
R}\right)\,,\nonumber\\
\mathcal{A}_{1^\prime 1^\prime
1^\prime}=&\,-\frac{9}{4}\left[2X_{uR}^3+X_{tR}^3+3X_{dR}^3\right.\nonumber\\
&\left.-X_{dR}\left(2X_{uR}^2+X_{tR}^2\right)-X_{dR}^2\left(2X_{uR}+X_{tR}\right)\right]\nonumber\\
&+\sum_{\alpha=e,\mu,\tau}\left(2X_{\alpha L}^3-X^3_{\alpha R}\right)\,.
\end{align}
Satisfying these conditions with rational charges is non-trivial and there is only one class of
solutions giving an anomaly-free model, up to lepton flavor permutations:
\begin{align}
\begin{split}
X_{uR}&=-X_{dR}-\frac{1}{3}X_{\mu R}\,,\quad X_{tR}=-4X_{dR}+\frac{2}{3}X_{\mu R}\,,\\
X_{e L}&=X_{dR}+\frac{1}{6}X_{\mu R}\,,\quad\hspace{6.5pt} X_{\mu
L}=-X_{dR}+\frac{5}{6}X_{\mu R}\,,\\
X_{\tau L}&=\frac{9}{2}X_{dR}-X_{\mu R}\,,\\
X_{e R}&=2X_{dR}+\frac{1}{3}X_{\mu R}\,,\quad\hspace{1.5pt} X_{\tau R}= 7 X_{dR}-\frac{4}{3}X_{\mu R}\,.
\end{split}
\end{align}
At this point we have two free charges in the above relations, \emph{i.e.} $X_{dR}$ and $X_{\mu R}$. 
Using $X_{\Phi_2}=0$, as discussed in Sec.~\ref{s:frame}, yields one relation between the two
charges. The remaining free charge, \emph{e.g.}
$X_{dR}$, is fixed by some normalization convention.

\section{Scalar potential}\label{app:scalar}
The Higgs doublets are parametrized as
%%%%
\be\label{eq:Phidoublets}
\Phi_i=  \begin{pmatrix}
\Phi_i^+\\
\Phi_i^0
\end{pmatrix}  =e^{i\theta_i}
\begin{pmatrix}
\varphi^+_i\\
\frac{1}{\sqrt{2}}\left(v_i+\rho_i+i\eta_i\right)
\end{pmatrix}\quad(i=1,2)\,.
\ee
%%%%
Their neutral components acquire vevs given by $\langle \Phi_j^0 \rangle
= e^{i \theta_j}   v_j/\sqrt{2} $ with  $v_i >0$ and  $(v_1^2 + v_2^2)^{1/2} \equiv v$.  In full
generality we can rotate away the phase of $\Phi_1$, leaving the second doublet with the phase
$\theta = \theta_2 - \theta_1$.    We parametrize the complex scalar singlet as
\be
S = e^{ -i  \alpha_{S}/2 }  \frac{(    v_S + R_0 + i I_0  )}{\sqrt{2}} \,,
\ee
with $v_S >0 $.  

Since we have in our model three scalar fields, $\{\Phi_1, \Phi_2, S\}$, the phase-blind
part of the scalar potential has a $\mathrm{U(1)}^3$ global invariance. In order to avoid
massless Goldstone bosons, we
need to charge the $S$ field in such a way that the phase-sensitive part breaks this symmetry down
to $\mathrm{U(1)}^2=\mathrm{U(1)}_Y\times \mathrm{U(1)}^\prime$. Gauge-invariant
combinations can be built only from $\Phi_1^\dagger\Phi_2$ and $S$, leaving us with the
possibilities
$\Phi_1^\dagger \Phi_2 S$, $\Phi_1^\dagger \Phi_2 S^2$ and combinations
involving complex conjugates.
For concreteness we choose $\Phi_1^\dagger \Phi_2 S^2$ to be invariant, which imposes 
$X_S=- 9/8$. The scalar potential then reads
\begin{align}
\begin{split}
V=&\,   m_i^2|\Phi_i|^2+\frac{\lambda_i}{2}|\Phi_i|^4 +\lambda_{3}|\Phi_1|^2|\Phi_2|^2\\
&+\lambda_{4}|\Phi_1^\dagger\Phi_2|^2+ \frac{b_2}{2} |S|^2+ \frac{d_2}{4} |S|^4+ \frac{\delta_{i}}{2}|\Phi_i|^2|S|^2\\
& - \frac{\delta_3}{4}  \left(\Phi_1^\dagger\Phi_2S^2 + \Phi_2^\dagger\Phi_1 (S^*)^2\right) \,.
\end{split}
\end{align}
All the parameters of the scalar potential but $\delta_3$ are real due to hermiticity.
The parameter $\delta_3$ has been chosen to be real and positive by rephasing the scalar field $S$
appropriately.

The vacuum expectation value of the potential is given by
\begin{align}
V_0 =&\, \frac{v^4}{16} \Bigl[  8 \frac{m_1^2}{v^2}  \cbt^2 + 2 \lambda_1 \cbt^4  + 4 \sbt^2 \left(2 \frac{m_2^2}{v^2} + (\lambda_3 + \lambda_4) \cbt^2\right)  \nonumber \\ &+  2 \lambda_2 \sbt^4 + 2 \hat v^2  \left(2 \frac{b_2}{v^2} + \delta_{1} \cbt^2 + \delta_{2} \sbt^2\right) + d_2\hat v^4  \nonumber \\&- 
   2\, \delta_3\, \cbt \sbt \hat v^2\cos(\alpha_S - \theta)  \Bigr] \,.
\end{align}
Here we have defined the ratio $\hat v = v_S/v $. The stability of the vacuum requires that
\be
0 = \frac{\partial V_0}{ \partial \theta } = -  \frac{\delta_3   v ^2 v_S^2}{16}  s_{2\beta}\sin(\alpha_S - \theta) \,.
\ee
By convention we choose $\alpha_S = \theta$, the other possibilities in $\alpha_S = \theta \mod
2\pi$ simply amount to an unphysical rephasing of the scalar field $S$. Stability also requires
$\partial V_0/\partial v_i  =  0$, to which the only non-trivial solution reads
\begin{align}
b_2 &= \frac{v^2}{2} \left[   - \delta_{1} \cbt^2 + \delta_3 \cbt \sbt - \delta_{2} \sbt^2  -d_2  \hat v^2   \right] \,, \nonumber \\
m_1^2 &=    \frac{v^2}{8} \left[    - 4 \left( \lambda_3  + \lambda_4  \right)  \sbt^2   - 4  \lambda_1 \cbt^2 + \delta_3 t_\beta \hat v^2   - 2 \delta_{1} \hat v^2   \right]    \,, \nonumber  \\
m_2^2 &=  \frac{v^2}{8 }  \left[   - 4 \left( \lambda_3 + \lambda_4\right)\cbt^2 - 4 \lambda_2 \sbt^2 + \delta_3 t_{\beta}^{-1} \hat v^2  - 2 \delta_{2} \hat v^2    \right] \,. 
\end{align}
The scalar potential is then determined in terms of 10 unknown parameters $\{v_S,\beta,
\lambda_{1-4}, d_2, \delta_{1-3}\}$, since $\theta=\alpha_S$ does not appear explicitly.

The masses of the physical CP-odd boson $A$ and the charged scalar are given by
\begin{align}
\begin{aligned}
M_{A}^2 &= \frac{ \delta_3 \, v_S^2}{4} \left( s_{2\beta}^{-1} + \dfrac{s_{2\beta}}{\hat v^2}       
\right)  \simeq   \frac{ \delta_3 \, v_S^2}{4 s_{2\beta}}     \,,\\
M_{H^{\pm}}^2  &= \frac{v_S^2}{4}  \left(     \frac{ \delta_3}{ s_{2\beta}  }  - \frac{2 \lambda_4}{
\hat v^2 }  \,   \right)   \simeq  \frac{ \delta_3  v_S^2}{4 s_{2\beta}  }\,,
\end{aligned}
\end{align}
respectively.
The physical states $H_{1,2,3}$ are expressed in terms of $\{ \rho_1, \rho_2, R_0 \}$ via an
orthogonal transformation,
\be
 \begin{pmatrix}
    H_1  \\  
    H_2 \\
    H_3
 \end{pmatrix}  = \mathcal{R}_S\,  \begin{pmatrix}
    \rho_1  \\  
      \rho_2 \\
    R_0
 \end{pmatrix}  \,.
\ee
The mass matrix for the CP-even bosons can be diagonalized analytically in a
perturbative expansion around $1/\hat v \ll 1$.  Including the leading corrections in $1/\hat v$ one obtains
\be
 \mathcal{R}_S \simeq  \left( \begin{array}{c c c }
    c_{\beta} &  s_{\beta} & \omega_{13}     \\
   - s_{\beta}   &  c_{\beta} & \omega_{23}   \\
   \omega_{23}s_\beta-\omega_{13}c_\beta &  -(\omega_{23}c_\beta+\omega_{13}s_\beta)  & 1     
   \end{array}\right)\, ,
\ee
with
\begin{align}
\begin{aligned}
\omega_{13}&=-\frac{2\delta_{1}c_\beta^2-\delta_3 s_{2\beta}+2\delta_{2}s_\beta^2}{2\hat{v}d_2} \,, \\
\omega_{23}&=\frac{\delta_3s_{4\beta}+2(\delta_{1}-\delta_{2})s^2_{2\beta}}{2\hat{v}(2d_2s_{2\beta}-\delta_3)}
\,. \\
\end{aligned}
\end{align}
The resulting masses for the CP-even scalars are given by
\begin{align}
M_{H_1}^2 \simeq&\, \frac{\tilde{\lambda}v^2}{2 d_2}\,,\quad M_{H_2}^2 \simeq \frac{\delta_3    v_S^2}{ 4 s_{2\beta} } \,, \quad M_{H_3}^2 \simeq  \dfrac{d_2 v_S^2}{2} \,,
\end{align}
with
\begin{align}
\begin{split}
\tilde{\lambda}=&\,  (  2 d_2 \lambda_1 - \delta_{1}^2   )   \cbt^4 + (\delta_{1} \cbt^2 + \delta_{2}    \sbt^2 )\delta_{3}s_{2\beta}     \\ 
&- \frac{2 \delta_{1}  \delta_{2}     + \delta_{3}^2}{4} s_{2\beta}^2  - (   \delta_{2}^2   - 2 d_2 \lambda_2  ) \sbt^4 \\
&+ d_2 ( \lambda_3 + \lambda_4) s_{2 \beta }^2         \,. \\
\end{split}
\end{align}

The exact expression for their 
Yukawa couplings in Eq.~\eqref{YukaLag} is
\begin{align}
\begin{aligned}
v \,Y_{f}^{H_k} =& \left[ c_{\beta} (\mathcal{R}_S)_{k1}   + s_{\beta}  (\mathcal{R}_S)_{k2}   \right]    D_{f}  \\ 
&+ \left[ s_{\beta} (\mathcal{R}_S)_{k1}   - c_{\beta}  (\mathcal{R}_S)_{k2}    \right]  N_{f} \,.
\end{aligned}
\end{align}

Finally, the vacuum solution in our models allow for complex vevs but that is not sufficient to
have spontaneous CP violation. In the weak gauge sector CP violation is manifest through the
invariant $\text{Tr}\left[H_d,H_u\right]^3$~\cite{Bernabeu:1986fc}. In the CP-invariant scenario,
\emph{i.e.} with real Yukawa matrices, the Hermitian combinations $H_{u,d}$ are given by
\begin{align}
\begin{split}
2H_u&=v_1^2\Delta_1\Delta_1^T+v_2^2\Delta_2\Delta_2^T+v_1v_2(\Delta_1\Delta_2^T+\Delta_2\Delta_1^T)c_\theta\\
&+iv_1v_2(\Delta_1\Delta_2^T-\Delta_2\Delta_1^T)s_\theta\,,\\
2H_d&=v_1^2\Gamma_1\Gamma_1^T+v_2^2\Gamma_2\Gamma_2^T+v_1v_2(\Gamma_2\Gamma_1^T+\Gamma_1\Gamma_2^T)c_\theta\\
&+iv_1v_2(\Gamma_2\Gamma_1^T-\Gamma_1\Gamma_2^T)s_\theta\,.
\end{split}
\end{align} 
In our model we get $\text{Tr}\left[H_d,H_u\right]^3=0$, implying the absence of CP violation in the gauge interactions when the only phase is carried by the scalar vev.   As a consequence, the source of CP violation for the weak currents in our model is present in the Yukawa couplings and will appear in the observables through the CKM mechanism. 

\end{appendix}


\begin{thebibliography}{99}

  

%\cite{Glashow:1961tr}
\bibitem{Glashow:1961tr} 
  S.~L.~Glashow,
  %``Partial Symmetries of Weak Interactions,''
  Nucl.\ Phys.\  {\bf 22}, 579 (1961).
  %%CITATION = NUPHA,22,579;%%
  %5773 citations counted in INSPIRE as of 24 Apr 2015
%
%\cite{Weinberg:1967tq}
%\bibitem{Weinberg:1967tq} 
  S.~Weinberg,
  %``A Model of Leptons,''
  Phys.\ Rev.\ Lett.\  {\bf 19}, 1264 (1967).
  %%CITATION = PRLTA,19,1264;%%
  %9389 citations counted in INSPIRE as of 24 Apr 2015
%
%\cite{Salam:1968rm}
%\bibitem{Salam:1968rm} 
  A.~Salam,
  %``Weak and Electromagnetic Interactions,''
  Conf.\ Proc.\ C {\bf 680519}, 367 (1968).
  %%CITATION = CONFP,C680519,367;%%
  %2515 citations counted in INSPIRE as of 24 Apr 2015

%\cite{Aaij:2014ora}
\bibitem{Aaij:2014ora}
  R.~Aaij {\it et al.}  [LHCb Collaboration],
  %``Test of lepton universality using $B^{+}\rightarrow K^{+}\ell^{+}\ell^{-}$ decays,''
  Phys.\ Rev.\ Lett.\  {\bf 113} (2014) 151601
  [arXiv:1406.6482 [hep-ex]].
  %%CITATION = ARXIV:1406.6482;%%
  %41 citations counted in INSPIRE as of 28 Mar 2015

  %\cite{Aaij:2013qta}
\bibitem{Aaij:2013qta}
  R.~Aaij {\it et al.}  [LHCb Collaboration],
  %``Measurement of Form-Factor-Independent Observables in the Decay $B^{0} \to K^{*0} \mu^+ \mu^-$,''
  Phys.\ Rev.\ Lett.\  {\bf 111} (2013) 191801
  [arXiv:1308.1707 [hep-ex]].
  %%CITATION = ARXIV:1308.1707;%%
  %117 citations counted in INSPIRE as of 28 mar 2015
LHCb-CONF-2015-002.
 
   %\cite{Altmannshofer:2014cfa}
\bibitem{Altmannshofer:2014cfa}
  W.~Altmannshofer, S.~Gori, M.~Pospelov and I.~Yavin,
  %``Quark flavor transitions in $L_\mu-L_\tau$ models,''
  Phys.\ Rev.\ D {\bf 89} (2014) 9,  095033
  [arXiv:1403.1269 [hep-ph]].
  %%CITATION = ARXIV:1403.1269;%%
  %33 citations counted in INSPIRE as of 21 Mar 2015
  
%\cite{Crivellin:2015mga}
\bibitem{Crivellin:2015mga}
  A.~Crivellin, G.~D'Ambrosio and J.~Heeck,
  %``Explaining $h\to\mu^\pm\tau^\mp$, $B\to K^* \mu^+\mu^-$ and $B\to K \mu^+\mu^-/B\to K e^+e^-$ in a two-Higgs-doublet model with gauged $L_\mu-L_\tau$,''
  Phys.\ Rev.\ Lett.\  {\bf 114} (2015) 151801
  [arXiv:1501.00993 [hep-ph]].
  %%CITATION = ARXIV:1501.00993;%%
  %22 citations counted in INSPIRE as of 24 Apr 2015
  
%\cite{Crivellin:2015lwa}
\bibitem{Crivellin:2015lwa}
  A.~Crivellin, G.~D'Ambrosio and J.~Heeck,
  %``Addressing the LHC flavor anomalies with horizontal gauge symmetries,''
  Phys.\ Rev.\ D {\bf 91} (2015) 7,  075006
  [arXiv:1503.03477 [hep-ph]].
  %%CITATION = ARXIV:1503.03477;%%
  %13 citations counted in INSPIRE as of 24 Apr 2015


  %\cite{Buras:2013dea}
\bibitem{Buras:2013dea}
  A.~J.~Buras, F.~De Fazio and J.~Girrbach,
  %``331 models facing new $b \to s\mu^+ \mu^-$ data,''
  JHEP {\bf 1402} (2014) 112
  [arXiv:1311.6729 [hep-ph]]
  %%CITATION = ARXIV:1311.6729;%%
  %51 citations counted in INSPIRE as of 31 Mar 2015
%
%\cite{Gauld:2013qja}
%\bibitem{Gauld:2013qja}
  R.~Gauld, F.~Goertz and U.~Haisch,
  %``An explicit Z'-boson explanation of the $B \to K^* \mu^+ \mu^-$ anomaly,''
  JHEP {\bf 1401} (2014) 069
  [arXiv:1310.1082 [hep-ph]].
  %%CITATION = ARXIV:1310.1082;%%
  %45 citations counted in INSPIRE as of 31 Mar 2015

  %\cite{Descotes-Genon:2013wba}
\bibitem{Descotes-Genon:2013wba}
  S.~Descotes-Genon, J.~Matias and J.~Virto,
  %``Understanding the $B \to K^*\mu^+\mu^-$ Anomaly,''
  Phys.\ Rev.\ D {\bf 88} (2013) 7,  074002
  [arXiv:1307.5683 [hep-ph]].
  %%CITATION = ARXIV:1307.5683;%%
  %98 citations counted in INSPIRE as of 23 Mar 2015
%
  %\cite{Altmannshofer:2013foa}
%\bibitem{Altmannshofer:2013foa}
  W.~Altmannshofer and D.~M.~Straub,
  %``New physics in $B \to K^*\mu\mu$?,''
  Eur.\ Phys.\ J.\ C {\bf 73} (2013) 2646
  [arXiv:1308.1501 [hep-ph]].
  %%CITATION = ARXIV:1308.1501;%%
  %88 citations counted in INSPIRE as of 31 Mar 2015
%
%\cite{Altmannshofer:2014rta}
%\bibitem{Altmannshofer:2014rta}
  W.~Altmannshofer and D.~M.~Straub,
  %``New physics in $b\to s$ transitions after LHC run 1,''
  arXiv:1411.3161 [hep-ph].
  %%CITATION = ARXIV:1411.3161;%%
  %35 citations counted in INSPIRE as of 24 Apr 2015
     
  %\cite{Crivellin:2015era}
\bibitem{Crivellin:2015era}
  A.~Crivellin {\it et al.},
  %``Lepton-flavour violating \boldmath$B$ decays in generic $Z^\prime$ models,''
  arXiv:1504.07928 [hep-ph].
  %%CITATION = ARXIV:1504.07928;%%

  %\cite{Sierra:2015fma}
\bibitem{Sierra:2015fma}
  D.~A.~Sierra, F.~Staub and A.~Vicente,
  %``Shedding light on the $b\to s$ anomalies with a dark sector,''
  arXiv:1503.06077 [hep-ph].
  %%CITATION = ARXIV:1503.06077;%%
%
%\cite{Buras:2013qja}
%\bibitem{Buras:2013qja}
  A.~J.~Buras and J.~Girrbach,
  %``Left-handed Z' and Z FCNC quark couplings facing new $b \to s \mu^+ \mu^-$ data,''
  JHEP {\bf 1312} (2013) 009
  [arXiv:1309.2466 [hep-ph]].
  %%CITATION = ARXIV:1309.2466;%%
  %52 citations counted in INSPIRE as of 31 Mar 2015
%
  %\cite{Gauld:2013qba}
%\bibitem{Gauld:2013qba}
  R.~Gauld, F.~Goertz and U.~Haisch,
  %``On minimal Z' explanations of the B->K*mu+mu- anomaly,''
  Phys.\ Rev.\ D {\bf 89} (2014) 1,  015005
  [arXiv:1308.1959 [hep-ph]].
  %%CITATION = ARXIV:1308.1959;%%
  %42 citations counted in INSPIRE as of 21 mar 2015

%\cite{Bhattacharya:2014wla}
\bibitem{Bhattacharya:2014wla}
  B.~Bhattacharya, A.~Datta, D.~London and S.~Shivashankara,
  %``Simultaneous Explanation of the $R_K$ and $R(D^{(*)})$ Puzzles,''
  Phys.\ Lett.\ B {\bf 742} (2015) 370
  [arXiv:1412.7164 [hep-ph]].
  %%CITATION = ARXIV:1412.7164;%%
  %5 citations counted in INSPIRE as of 23 mar 2015

%\cite{Varzielas:2015iva}
\bibitem{Varzielas:2015iva}
  I.~de Medeiros Varzielas and G.~Hiller,
  %``Clues for flavor from rare lepton and quark decays,''
  JHEP {\bf 1506} (2015) 072
  [arXiv:1503.01084 [hep-ph]].
  %%CITATION = ARXIV:1503.01084;%%
  %12 citations counted in INSPIRE as of 26 Jun 2015
  
  %\cite{Hiller:2014yaa}
\bibitem{Hiller:2014yaa}
  G.~Hiller and M.~Schmaltz,
  %``$R_K$ and future $b \to s \ell \ell$ physics beyond the standard model opportunities,''
  Phys.\ Rev.\ D {\bf 90} (2014) 5,  054014
  [arXiv:1408.1627 [hep-ph]].
  %%CITATION = ARXIV:1408.1627;%%
  %22 citations counted in INSPIRE as of 21 Mar 2015

%\cite{Gripaios:2014tna}
\bibitem{Gripaios:2014tna}
  B.~Gripaios, M.~Nardecchia and S.~A.~Renner,
  %``Composite leptoquarks and anomalies in $B$-meson decays,''
  JHEP {\bf 1505} (2015) 006
  [arXiv:1412.1791 [hep-ph]].
  %%CITATION = ARXIV:1412.1791;%%
  %12 citations counted in INSPIRE as of 24 Apr 2015
%
  %\cite{Niehoff:2015bfa}
%\bibitem{Niehoff:2015bfa}
  C.~Niehoff, P.~Stangl and D.~M.~Straub,
  %``Violation of lepton flavour universality in composite Higgs models,''
  arXiv:1503.03865 [hep-ph].
  %%CITATION = ARXIV:1503.03865;%%
  %1 citations counted in INSPIRE as of 23 Mar 2015
%
%\cite{Becirevic:2015asa}
%\bibitem{Becirevic:2015asa}
  D.~Becirevic, S.~Fajfer and N.~Kosnik,
  %``Lepton flavor non-universality in $b \to s \ell^+ \ell^-$ processes,''
  arXiv:1503.09024 [hep-ph].
  %%CITATION = ARXIV:1503.09024;%%


  %\cite{Langacker:2008yv}
\bibitem{Langacker:2008yv}
 For a review on $Z^{\prime}$ models see: P.~Langacker,
  %``The Physics of Heavy $Z^\prime$ Gauge Bosons,''
  Rev.\ Mod.\ Phys.\  {\bf 81} (2009) 1199
  [arXiv:0801.1345 [hep-ph]].
  %%CITATION = ARXIV:0801.1345;%%
  %590 citations counted in INSPIRE as of 21 Mar 2015
 


%\cite{Langacker:2000ju}
\bibitem{Langacker:2000ju} 
  P.~Langacker and M.~Plumacher,
  %``Flavor changing effects in theories with a heavy $Z^\prime$ boson with family nonuniversal couplings,''
  Phys.\ Rev.\ D {\bf 62}, 013006 (2000)
  [hep-ph/0001204].
  %%CITATION = HEP-PH/0001204;%%
  %231 citations counted in INSPIRE as of 14 Apr 2015
%
  %\cite{Barger:2003hg}
%\bibitem{Barger:2003hg}
  V.~Barger, C.~W.~Chiang, P.~Langacker and H.~S.~Lee,
  %``$Z^\prime$ mediated flavor changing neutral currents in $B$ meson decays,''
  Phys.\ Lett.\ B {\bf 580} (2004) 186
  [hep-ph/0310073].
  %%CITATION = HEP-PH/0310073;%%
  %120 citations counted in INSPIRE as of 24 Apr 2015
%
  %\cite{He:2004it}
%\bibitem{He:2004it}
  X.~G.~He and G.~Valencia,
  %``K+ ---> pi+ nu anti-nu and FCNC from non-universal Z' bosons,''
  Phys.\ Rev.\ D {\bf 70} (2004) 053003
  [hep-ph/0404229].
  %%CITATION = HEP-PH/0404229;%%
  %36 citations counted in INSPIRE as of 24 Apr 2015
%
  %\cite{Chen:2006vs}
%\bibitem{Chen:2006vs}
  C.~H.~Chen and H.~Hatanaka,
  %``Nonuniversal Z-prime couplings in B decays,''
  Phys.\ Rev.\ D {\bf 73} (2006) 075003
  [hep-ph/0602140].
  %%CITATION = HEP-PH/0602140;%%
  %35 citations counted in INSPIRE as of 24 Apr 2015
%
  %\cite{Chang:2010zy}
%\bibitem{Chang:2010zy}
  Q.~Chang, X.~Q.~Li and Y.~D.~Yang,
  %``$B \to K^{\ast} l^+ l^-$, $K l^+ l^-$ decays in a family non-universal $Z^{\prime}$ model,''
  JHEP {\bf 1004} (2010) 052
  [arXiv:1002.2758 [hep-ph]].
  %%CITATION = ARXIV:1002.2758;%%
  %29 citations counted in INSPIRE as of 24 Apr 2015
%
  %\cite{Li:2012xc}
%\bibitem{Li:2012xc}
  X.~Q.~Li, Y.~M.~Li, G.~R.~Lu and F.~Su,
  %``$B_s^0-\bar B_s^0$ mixing in a family non-universal $Z^{\prime}$ model revisited,''
  JHEP {\bf 1205} (2012) 049
  [arXiv:1204.5250 [hep-ph]].
  %%CITATION = ARXIV:1204.5250;%%
  %9 citations counted in INSPIRE as of 24 Apr 2015
%
%\cite{Chang:2013hba}
%\bibitem{Chang:2013hba}
  Q.~Chang, X.~Q.~Li and Y.~D.~Yang,
  %``A comprehensive analysis of hadronic b $\to$ s transitions in a family non-universal Z-prime model,''
  J.\ Phys.\ G {\bf 41} (2014) 105002
  [arXiv:1312.1302 [hep-ph]].
  %%CITATION = ARXIV:1312.1302;%%
  %4 citations counted in INSPIRE as of 24 Apr 2015

    %\cite{Buras:2012jb}
\bibitem{Buras:2012jb}
  A.~J.~Buras, F.~De Fazio and J.~Girrbach,
  %``The Anatomy of Z' and Z with Flavour Changing Neutral Currents in the Flavour Precision Era,''
  JHEP {\bf 1302} (2013) 116
  [arXiv:1211.1896 [hep-ph]].
  %%CITATION = ARXIV:1211.1896;%%
  %53 citations counted in INSPIRE as of 31 Mar 2015

  
%\cite{He:1990pn}
\bibitem{He:1990pn} 
  X.~G.~He, G.~C.~Joshi, H.~Lew and R.~R.~Volkas,
  %``NEW Z-prime PHENOMENOLOGY,''
  Phys.\ Rev.\ D {\bf 43}, 22 (1991);
  X.~G.~He, G.~C.~Joshi, H.~Lew and R.~R.~Volkas,
  %``Simplest Z-prime model,''
  Phys.\ Rev.\ D {\bf 44}, 2118 (1991).  


  
  %\cite{Binetruy:1996cs}
\bibitem{Binetruy:1996cs}
  P.~Binetruy, S.~Lavignac, S.~T.~Petcov and P.~Ramond,
  %``Quasidegenerate neutrinos from an Abelian family symmetry,''
  Nucl.\ Phys.\ B {\bf 496} (1997) 3
  [hep-ph/9610481].
  %%CITATION = HEP-PH/9610481;%%
  %106 citations counted in INSPIRE as of 31 Mar 2015
%
  %\cite{Bell:2000vh}
%\bibitem{Bell:2000vh}
  N.~F.~Bell and R.~R.~Volkas,
  %``Bottom up model for maximal muon-neutrino - tau-neutrino mixing,''
  Phys.\ Rev.\ D {\bf 63} (2001) 013006
  [hep-ph/0008177].
  %%CITATION = HEP-PH/0008177;%%
  %15 citations counted in INSPIRE as of 31 Mar 2015
%
  %\cite{Choubey:2004hn}
%\bibitem{Choubey:2004hn}
  S.~Choubey and W.~Rodejohann,
  %``A Flavor symmetry for quasi-degenerate neutrinos: L(mu) - L(tau),''
  Eur.\ Phys.\ J.\ C {\bf 40} (2005) 259
  [hep-ph/0411190].
  %%CITATION = HEP-PH/0411190;%%
  %70 citations counted in INSPIRE as of 31 Mar 2015
%
  %\cite{Heeck:2011wj}
%\bibitem{Heeck:2011wj}
  J.~Heeck and W.~Rodejohann,
  %``Gauged L_mu - L_tau Symmetry at the Electroweak Scale,''
  Phys.\ Rev.\ D {\bf 84} (2011) 075007
  [arXiv:1107.5238 [hep-ph]].
  %%CITATION = ARXIV:1107.5238;%%
  %26 citations counted in INSPIRE as of 31 Mar 2015
   %\cite{Cebola:2013hta}
%\bibitem{Cebola:2013hta}
  L.~M.~Cebola, D.~Emmanuel-Costa and R.~Gonzalez Felipe,
  %``Anomaly-free U(1) gauge symmetries in neutrino seesaw flavor models,''
  Phys.\ Rev.\ D {\bf 88} (2013) 11,  116008
  [arXiv:1309.1709 [hep-ph]].
  %%CITATION = ARXIV:1309.1709;%%
  %2 citations counted in INSPIRE as of 31 Mar 2015

 %\cite{Leurer:1992wg}
\bibitem{Leurer:1992wg}
  M.~Leurer, Y.~Nir and N.~Seiberg,
  %``Mass matrix models,''
  Nucl.\ Phys.\ B {\bf 398} (1993) 319
  [hep-ph/9212278].
  %%CITATION = HEP-PH/9212278;%%
  %297 citations counted in INSPIRE as of 30 Mar 2015
%\cite{Felipe:2014zka}
%\bibitem{Felipe:2014zka}
  R.~Gonzalez Felipe, I.~P.~Ivanov, C.~C.~Nishi, H.~Serodio and J.~P.~Silva,
  %``Constraining multi-Higgs flavour models,''
  Eur.\ Phys.\ J.\ C {\bf 74} (2014) 7,  2953
  [arXiv:1401.5807 [hep-ph]].
  %%CITATION = ARXIV:1401.5807;%%
  %3 citations counted in INSPIRE as of 30 mar 2015       
  
  

  
  
  %\cite{Cabibbo:1963yz}
\bibitem{Cabibbo:1963yz}
  N.~Cabibbo,
  %``Unitary Symmetry and Leptonic Decays,''
  Phys.\ Rev.\ Lett.\  {\bf 10} (1963) 531;
  %%CITATION = PRLTA,10,531;%%
  %4456 citations counted in INSPIRE as of 19 May 2014
  %\cite{Kobayashi:1973fv}
%\bibitem{Kobayashi:1973fv}
  M.~Kobayashi and T.~Maskawa,
  %``CP Violation in the Renormalizable Theory of Weak Interaction,''
  Prog.\ Theor.\ Phys.\  {\bf 49} (1973) 652.
  %%CITATION = PTPKA,49,652;%%
  %7690 citations counted in INSPIRE as of 19 May 2014
  

    
   

%\cite{Branco:1996bq}
\bibitem{Branco:1996bq}
  G.~C.~Branco, W.~Grimus and L.~Lavoura,
  %``Relating the scalar flavor changing neutral couplings to the CKM matrix,''
  Phys.\ Lett.\ B {\bf 380} (1996) 119
  [hep-ph/9601383].
  %%CITATION = HEP-PH/9601383;%%
  %60 citations counted in INSPIRE as of 21 mar 2015
  

  
      
  %\cite{Glashow:1976nt}
\bibitem{Glashow:1976nt}
  S.~L.~Glashow and S.~Weinberg,
  %``Natural Conservation Laws for Neutral Currents,''
  Phys.\ Rev.\ D {\bf 15} (1977) 1958;
  %%CITATION = PHRVA,D15,1958;%%
  %1258 citations counted in INSPIRE as of 22 Jul 2014
  %\cite{Paschos:1976ay}
%\bibitem{Paschos:1976ay}
  E.~A.~Paschos,
  %``Diagonal Neutral Currents,''
  Phys.\ Rev.\ D {\bf 15} (1977) 1966.
  %%CITATION = PHRVA,D15,1966;%%
  %341 citations counted in INSPIRE as of 22 Jul 2014


    %\cite{Botella:2011ne}
\bibitem{Botella:2011ne}
  F.~J.~Botella, G.~C.~Branco, M.~Nebot and M.~N.~Rebelo,
  %``Two-Higgs Leptonic Minimal Flavour Violation,''
  JHEP {\bf 1110} (2011) 037
  [arXiv:1102.0520 [hep-ph]].
  %%CITATION = ARXIV:1102.0520;%%
  %14 citations counted in INSPIRE as of 21 mar 2015

  
 %\cite{Botella:2014ska}
\bibitem{Botella:2014ska}
  F.~J.~Botella {\it et al.},
  %``Physical Constraints on a Class of Two-Higgs Doublet Models with FCNC at tree level,''
  JHEP {\bf 1407} (2014) 078
  [arXiv:1401.6147 [hep-ph]].
  %%CITATION = ARXIV:1401.6147;%%
  %12 citations counted in INSPIRE as of 24 Mar 2015 
 %\cite{Bhattacharyya:2014nja}
%\bibitem{Bhattacharyya:2014nja}
  G.~Bhattacharyya, D.~Das and A.~Kundu,
  %``Feasibility of light scalars in a class of two-Higgs-doublet models and their decay signatures,''
  Phys.\ Rev.\ D {\bf 89} (2014) 9,  095029
  [arXiv:1402.0364 [hep-ph]].
  %%CITATION = ARXIV:1402.0364;%%
  %8 citations counted in INSPIRE as of 24 Mar 2015 
  
  
  
  %\cite{Celis:2014iua}
\bibitem{Celis:2014iua}
  A.~Celis, J.~Fuentes-Martin and H.~Serodio,
  %``An invisible axion model with controlled FCNCs at tree level,''
  Phys.\ Lett.\ B {\bf 741} (2015) 117
  [arXiv:1410.6217 [hep-ph].
  %%CITATION = ARXIV:1410.6217;%%
  %1 citations counted in INSPIRE as of 28 mar 2015
  %\cite{Celis:2014jua}
%\bibitem{Celis:2014jua}
 % A.~Celis, J.~Fuentes-Martin and H.~Serodio,
  %``A class of invisible axion models with FCNCs at tree level,''
  JHEP {\bf 1412} (2014) 167
  [arXiv:1410.6218 [hep-ph]].
  %%CITATION = ARXIV:1410.6218;%%
  %2 citations counted in INSPIRE as of 28 Mar 2015
  
    
  %\cite{Green:1984sg}
\bibitem{Green:1984sg}
  M.~B.~Green and J.~H.~Schwarz,
  %``Anomaly Cancellation in Supersymmetric D=10 Gauge Theory and Superstring Theory,''
  Phys.\ Lett.\ B {\bf 149} (1984) 117;
  %%CITATION = PHLTA,B149,117;%%
  %2304 citations counted in INSPIRE as of 28 Mar 2015
  %\cite{Green:1984qs}
  %\bibitem{Green:1984qs}
  M.~B.~Green and J.~H.~Schwarz,
  %``The Hexagon Gauge Anomaly in Type I Superstring Theory,''
  Nucl.\ Phys.\ B {\bf 255} (1985) 93;
  %%CITATION = NUPHA,B255,93;%%
  %226 citations counted in INSPIRE as of 28 Mar 2015
   %\cite{Green:1984bx}
%\bibitem{Green:1984bx}
  M.~B.~Green, J.~H.~Schwarz and P.~C.~West,
  %``Anomaly Free Chiral Theories in Six-Dimensions,''
  Nucl.\ Phys.\ B {\bf 254} (1985) 327.
  %%CITATION = NUPHA,B254,327;%%
  %230 citations counted in INSPIRE as of 28 Mar 2015
  

  %\cite{Glashow:2014iga}
\bibitem{Glashow:2014iga}
  S.~L.~Glashow, D.~Guadagnoli and K.~Lane,
  %``Lepton Flavor Violation in $B$ Decays?,''
  Phys.\ Rev.\ Lett.\  {\bf 114} (2015) 091801
  [arXiv:1411.0565 [hep-ph]].
  %%CITATION = ARXIV:1411.0565;%%
  %18 citations counted in INSPIRE as of 12 Apr 2015
  
  %\cite{Boucenna:2015raa}
\bibitem{Boucenna:2015raa}
  S.~M.~Boucenna, J.~W.~F.~Valle and A.~Vicente,
  %``Are the B decay anomalies related to neutrino oscillations?,''
  arXiv:1503.07099 [hep-ph].
  %%CITATION = ARXIV:1503.07099;%%
  %2 citations counted in INSPIRE as of 12 Apr 2015  
  
  %\cite{Babu:1997st}
\bibitem{Babu:1997st}
  K.~S.~Babu, C.~F.~Kolda and J.~March-Russell,
  %``Implications of generalized Z - Z-prime mixing,''
  Phys.\ Rev.\ D {\bf 57} (1998) 6788
  [hep-ph/9710441].
  %%CITATION = HEP-PH/9710441;%%
  %149 citations counted in INSPIRE as of 21 Mar 2015
  
    
    %\cite{Agashe:2014kda}
\bibitem{Agashe:2014kda}
  K.~A.~Olive {\it et al.}  [Particle Data Group Collaboration],
  %``Review of Particle Physics,''
  Chin.\ Phys.\ C {\bf 38} (2014) 090001.
  %%CITATION = CHPHD,C38,090001;%%
  %811 citations counted in INSPIRE as of 28 mar 2015
  
  %\cite{Erler:1999ub}
\bibitem{Erler:1999ub}
  J.~Erler and P.~Langacker,
  %``Constraints on extended neutral gauge structures,''
  Phys.\ Lett.\ B {\bf 456} (1999) 68
  [hep-ph/9903476].
  %%CITATION = HEP-PH/9903476;%%
  %112 citations counted in INSPIRE as of 12 Apr 2015
  
  
  
  
  %\cite{Foot:2013hna}
\bibitem{Foot:2013hna}
  R.~Foot, A.~Kobakhidze, K.~L.~McDonald and R.~R.~Volkas,
  %``Poincare protection for a natural electroweak scale,''
  Phys.\ Rev.\ D {\bf 89} (2014) 11,  115018
  [arXiv:1310.0223 [hep-ph]].
  %%CITATION = ARXIV:1310.0223;%%
  %25 citations counted in INSPIRE as of 27 Mar 2015
  
  
  



  
%\cite{Amhis:2014hma}
\bibitem{Amhis:2014hma} 
  Y.~Amhis {\it et al.}  [Heavy Flavor Averaging Group (HFAG) Collaboration],
  %``Averages of $b$-hadron, $c$-hadron, and $\tau$-lepton properties as of summer 2014,''
  arXiv:1412.7515 [hep-ex]. Online update available at
  \texttt{http://www.slac.stanford.edu/xorg/hfag/} .
  %%CITATION = ARXIV:1412.7515;%%
  %33 citations counted in INSPIRE as of 17 Apr 2015
  
  

\bibitem{LatticeProgress}
%\cite{Bouchard:2014eea}
%\bibitem{Bouchard:2014eea} 
  C.~M.~Bouchard {\it et al.},
  %``Neutral B-meson mixing parameters in and beyond the SM with 2+1 flavor lattice QCD,''
  arXiv:1412.5097 [hep-lat.
  %%CITATION = ARXIV:1412.5097;%%
  %2 citations counted in INSPIRE as of 17 Apr 2015
%\cite{Dowdall:2014qka}
%\bibitem{Dowdall:2014qka} 
  R.~J.~Dowdall {\it et al.},
  %``B-meson mixing from full lattice QCD with physical u, d, s and c quarks,''
  arXiv:1411.6989 [hep-lat.
  %%CITATION = ARXIV:1411.6989;%%
  %2 citations counted in INSPIRE as of 17 Apr 2015
%\cite{Aoki:2014nga}
%\bibitem{Aoki:2014nga} 
  Y.~Aoki, T.~Ishikawa, T.~Izubuchi, C.~Lehner and A.~Soni,
  %``Neutral $B$ meson mixings and $B$ meson decay constants with static heavy and domain-wall light quarks,''
  arXiv:1406.6192 [hep-lat].
  %%CITATION = ARXIV:1406.6192;%%
  %6 citations counted in INSPIRE as of 17 Apr 2015


%\cite{Aoki:2013ldr}
\bibitem{Aoki:2013ldr} 
  S.~Aoki {\it et al.},
  %``Review of lattice results concerning low-energy particle physics,''
  Eur.\ Phys.\ J.\ C {\bf 74}, 2890 (2014)
  [arXiv:1310.8555 [hep-lat]]. See also \texttt{http://itpwiki.unibe.ch/flag/index.php} .
  %%CITATION = ARXIV:1310.8555;%%
  %184 citations counted in INSPIRE as of 17 Apr 2015
  
  

  
  %\cite{Charles:2004jd}
\bibitem{Charles:2004jd}
  J.~Charles {\it et al.}  [CKMfitter Group Collaboration],
  %``CP violation and the CKM matrix: Assessing the impact of the asymmetric $B$ factories,''
  Eur.\ Phys.\ J.\ C {\bf 41} (2005) 1
  [hep-ph/0406184]. Updated results and plots available at: {\tt http://ckmfitter.in2p3.fr}.
  %%CITATION = HEP-PH/0406184;%%


  
    %\cite{Buras:2013uqa}
\bibitem{Buras:2013uqa}
  A.~J.~Buras, R.~Fleischer, J.~Girrbach and R.~Knegjens,
  %``Probing New Physics with the B_s to {\mu}+ {\mu}- Time-Dependent Rate,''
  JHEP {\bf 1307} (2013) 77
  [arXiv:1303.3820 [hep-ph]].
  %%CITATION = ARXIV:1303.3820;%%
  %66 citations counted in INSPIRE as of 16 Apr 2015
%
  %\cite{Buras:2013rqa}
%\bibitem{Buras:2013rqa}
  A.~J.~Buras, F.~De Fazio, J.~Girrbach, R.~Knegjens and M.~Nagai,
  %``The Anatomy of Neutral Scalars with FCNCs in the Flavour Precision Era,''
  JHEP {\bf 1306} (2013) 111
  [arXiv:1303.3723 [hep-ph]].
  %%CITATION = ARXIV:1303.3723;%%
  %16 citations counted in INSPIRE as of 16 Apr 2015

      
\bibitem{Ciuchini:2000de}
  M.~Ciuchini {\it et al.},
  %``2000 CKM triangle analysis: A Critical review with updated experimental inputs and theoretical parameters,''
  JHEP {\bf 0107} (2001) 013
  [hep-ph/0012308]. Updated results and plots available at: {\tt http://www.utfit.org}.
  %%CITATION = HEP-PH/0012308;%%

 %\cite{Altmannshofer:2014pba}
\bibitem{Altmannshofer:2014pba}
  W.~Altmannshofer, S.~Gori, M.~Pospelov and I.~Yavin,
  %``Neutrino Trident Production: A Powerful Probe of New Physics with Neutrino Beams,''
  Phys.\ Rev.\ Lett.\  {\bf 113} (2014) 091801
  [arXiv:1406.2332 [hep-ph]].
  %%CITATION = ARXIV:1406.2332;%%
  %13 citations counted in INSPIRE as of 29 Mar 2015
  

  %\cite{Geiregat:1990gz}
\bibitem{Geiregat:1990gz}
  D.~Geiregat {\it et al.}  [CHARM-II Collaboration],
  %``First observation of neutrino trident production,''
  Phys.\ Lett.\ B {\bf 245} (1990) 271.
  %%CITATION = PHLTA,B245,271;%%
  %22 citations counted in INSPIRE as of 31 Mar 2015
%
  %\cite{Mishra:1991bv}
%\bibitem{Mishra:1991bv}
  S.~R.~Mishra {\it et al.}  [CCFR Collaboration],
  %``Neutrino tridents and W Z interference,''
  Phys.\ Rev.\ Lett.\  {\bf 66} (1991) 3117.
  %%CITATION = PRLTA,66,3117;%%
  %25 citations counted in INSPIRE as of 31 Mar 2015
%
  %\cite{Adams:1999mn}
%\bibitem{Adams:1999mn}
  T.~Adams {\it et al.}  [NuTeV Collaboration],
  %``Evidence for diffractive charm production in muon-neutrino Fe and anti-muon-neutrino Fe scattering at the Tevatron,''
  Phys.\ Rev.\ D {\bf 61} (2000) 092001
  [hep-ex/9909041].
  %%CITATION = HEP-EX/9909041;%%
  %27 citations counted in INSPIRE as of 31 Mar 2015

%\cite{Wood:1997zq}
\bibitem{Wood:1997zq}
  C.~S.~Wood {\it et al.},
  %``Measurement of parity nonconservation and an anapole moment in cesium,''
  Science {\bf 275} (1997) 1759.
  %%CITATION = SCIEA,275,1759;%%
  %428 citations counted in INSPIRE as of 24 Apr 2015
%
%\cite{Bennett:1999pd}
%\bibitem{Bennett:1999pd}
  S.~C.~Bennett and C.~E.~Wieman,
  %``Measurement of the 6S ---> 7S transition polarizability in atomic cesium and an improved test of the Standard Model,''
  Phys.\ Rev.\ Lett.\  {\bf 82} (1999) 2484
   [Phys.\ Rev.\ Lett.\  {\bf 83} (1999) 889]
  [hep-ex/9903022].
  %%CITATION = HEP-EX/9903022;%%
  %297 citations counted in INSPIRE as of 24 Apr 2015
%
  %\cite{Derevianko:2000dt}
%\bibitem{Derevianko:2000dt}
  A.~Derevianko,
  %``Reconciliation of the measurement of parity nonconservation in Cs with the standard model,''
  Phys.\ Rev.\ Lett.\  {\bf 85} (2000) 1618
  [hep-ph/0005274].
  %%CITATION = HEP-PH/0005274;%%
  %70 citations counted in INSPIRE as of 24 Apr 2015
%
  %\cite{Carpentier:2010ue}
%\bibitem{Carpentier:2010ue}
  M.~Carpentier and S.~Davidson,
  %``Constraints on two-lepton, two quark operators,''
  Eur.\ Phys.\ J.\ C {\bf 70} (2010) 1071
  [arXiv:1008.0280 [hep-ph]].
  %%CITATION = ARXIV:1008.0280;%%
  %30 citations counted in INSPIRE as of 24 Apr 2015
%    
%\cite{Diener:2011jt}
%\bibitem{Diener:2011jt}
  R.~Diener, S.~Godfrey and I.~Turan,
  %``Constraining Extra Neutral Gauge Bosons with Atomic Parity Violation Measurements,''
  Phys.\ Rev.\ D {\bf 86} (2012) 115017
  [arXiv:1111.4566 [hep-ph]].
  %%CITATION = ARXIV:1111.4566;%%
  %12 citations counted in INSPIRE as of 24 Apr 2015
%
  %\cite{Dzuba:2012kx}
%\bibitem{Dzuba:2012kx}
  V.~A.~Dzuba, J.~C.~Berengut, V.~V.~Flambaum and B.~Roberts,
  %``Revisiting parity non-conservation in cesium,''
  Phys.\ Rev.\ Lett.\  {\bf 109} (2012) 203003
  [arXiv:1207.5864 [hep-ph]].
  %%CITATION = ARXIV:1207.5864;%%
  %43 citations counted in INSPIRE as of 24 Apr 2015
%
  %\cite{Roberts:2014bka}
%\bibitem{Roberts:2014bka}
  B.~M.~Roberts, V.~A.~Dzuba and V.~V.~Flambaum,
  %``Parity and Time-Reversal Violation in Atomic Systems,''
  arXiv:1412.6644 [physics.atom-ph].
  %%CITATION = ARXIV:1412.6644;%%
  %1 citations counted in INSPIRE as of 24 Apr 2015

%\cite{Botella:2012ab}
\bibitem{Botella:2012ab} 
  F.~J.~Botella, G.~C.~Branco and M.~N.~Rebelo,
  %``Invariants and Flavour in the General Two-Higgs Doublet Model,''
  Phys.\ Lett.\ B {\bf 722}, 76 (2013)
  [arXiv:1210.8163 [hep-ph]].
  %%CITATION = ARXIV:1210.8163;%%
  %7 citations counted in INSPIRE as of 17 Apr 2015
%
%\cite{Jung:2013hka}
%\bibitem{Jung:2013hka}
  M.~Jung and A.~Pich,
  %``Electric Dipole Moments in Two-Higgs-Doublet Models,''
  JHEP {\bf 1404} (2014) 076
  [arXiv:1308.6283 [hep-ph]].
  %%CITATION = ARXIV:1308.6283;%%
  %24 citations counted in INSPIRE as of 18 Apr 2015  
  
%\cite{Crivellin:2013wna}
\bibitem{Crivellin:2013wna} 
  A.~Crivellin, A.~Kokulu and C.~Greub,
  %``Flavor-phenomenology of two-Higgs-doublet models with generic Yukawa structure,''
  Phys.\ Rev.\ D {\bf 87}, no. 9, 094031 (2013)
  [arXiv:1303.5877 [hep-ph]].
  %%CITATION = ARXIV:1303.5877;%%
  %51 citations counted in INSPIRE as of 17 Apr 2015

%\cite{Chiang:2006we}
\bibitem{Chiang:2006we} 
  C.~W.~Chiang, N.~G.~Deshpande and J.~Jiang,
  %``Flavor changing effects in family nonuniversal $Z^\prime$ models,''
  JHEP {\bf 0608}, 075 (2006)
  [hep-ph/0606122].
  %%CITATION = HEP-PH/0606122;%%
  %41 citations counted in INSPIRE as of 17 Apr 2015


%\cite{Jegerlehner:2009ry}
\bibitem{Jegerlehner:2009ry} 
  F.~Jegerlehner and A.~Nyffeler,
  %``The Muon g-2,''
  Phys.\ Rept.\  {\bf 477}, 1 (2009)
  [arXiv:0902.3360 [hep-ph].
  %%CITATION = ARXIV:0902.3360;%%
  %432 citations counted in INSPIRE as of 17 Apr 2015
%\cite{Queiroz:2014zfa}
%\bibitem{Queiroz:2014zfa} 
  F.~S.~Queiroz and W.~Shepherd,
  %``New Physics Contributions to the Muon Anomalous Magnetic Moment: A Numerical Code,''
  Phys.\ Rev.\ D {\bf 89}, no. 9, 095024 (2014)
  [arXiv:1403.2309 [hep-ph]].
  %%CITATION = ARXIV:1403.2309;%%
  %17 citations counted in INSPIRE as of 17 Apr 2015

      %\cite{Aad:2012hf}
\bibitem{Aad:2012hf}
  G.~Aad {\it et al.}  [ATLAS Collaboration],
  %``Search for high-mass resonances decaying to dilepton final states in pp collisions at s**(1/2) = 7-TeV with the ATLAS detector,''
  JHEP {\bf 1211} (2012) 138
  [arXiv:1209.2535 [hep-ex]].
  %%CITATION = ARXIV:1209.2535;%%
  %83 citations counted in INSPIRE as of 12 Apr 2015
%
  %\cite{Aad:2014cka}
%\bibitem{Aad:2014cka}
  G.~Aad {\it et al.}  [ATLAS Collaboration],
  %``Search for high-mass dilepton resonances in pp collisions at $\sqrt{s}=8$??TeV with the ATLAS detector,''
  Phys.\ Rev.\ D {\bf 90} (2014) 5,  052005
  [arXiv:1405.4123 [hep-ex]].
  %%CITATION = ARXIV:1405.4123;%%
  %53 citations counted in INSPIRE as of 12 Apr 2015

  %\cite{Chatrchyan:2012it}
\bibitem{Chatrchyan:2012it}
  S.~Chatrchyan {\it et al.}  [CMS Collaboration],
  %``Search for narrow resonances in dilepton mass spectra in $pp$ collisions at $\sqrt{s}=7$ TeV,''
  Phys.\ Lett.\ B {\bf 714} (2012) 158
  [arXiv:1206.1849 [hep-ex]].
  %%CITATION = ARXIV:1206.1849;%%
  %97 citations counted in INSPIRE as of 12 Apr 2015
%
  %\cite{Khachatryan:2014fba}
%\bibitem{Khachatryan:2014fba}
  V.~Khachatryan {\it et al.}  [CMS Collaboration],
  %``Search for physics beyond the standard model in dilepton mass spectra in proton-proton collisions at $ \sqrt{s}=8 $ TeV,''
  JHEP {\bf 1504} (2015) 025
  [arXiv:1412.6302 [hep-ex]].
  %%CITATION = ARXIV:1412.6302;%%
  %11 citations counted in INSPIRE as of 24 Apr 2015


%\cite{Accomando:2013sfa}
\bibitem{Accomando:2013sfa}
  E.~Accomando {\it et al.},
  %``Z' at the LHC: Interference and Finite Width Effects in Drell-Yan,''
  JHEP {\bf 1310} (2013) 153
  [arXiv:1304.6700 [hep-ph]].
  %%CITATION = ARXIV:1304.6700;%%
  %15 citations counted in INSPIRE as of 24 Apr 2015

  
  %\cite{Carena:2004xs}
\bibitem{Carena:2004xs}
  M.~Carena, A.~Daleo, B.~A.~Dobrescu and T.~M.~P.~Tait,
  %``$Z^\prime$ gauge bosons at the Tevatron,''
  Phys.\ Rev.\ D {\bf 70} (2004) 093009
  [hep-ph/0408098].
  %%CITATION = HEP-PH/0408098;%%
  %303 citations counted in INSPIRE as of 31 Mar 2015
%
\bibitem{Accomando:2010fz}
%\bibitem{Accomando:2010fz}
  E.~Accomando  {\it et al.},
  %``Z' physics with early LHC data,''
  Phys.\ Rev.\ D {\bf 83} (2011) 075012
  [arXiv:1010.6058 [hep-ph]].
  %%CITATION = ARXIV:1010.6058;%%
  %74 citations counted in INSPIRE as of 11 Apr 2015
  
  
  %\cite{Aad:2014wca}
\bibitem{Aad:2014wca}
  G.~Aad {\it et al.}  [ATLAS Collaboration],
  %``Search for contact interactions and large extra dimensions in the dilepton channel using proton-proton collisions at $\sqrt{s}$ = 8 TeV with the ATLAS detector,''
  Eur.\ Phys.\ J.\ C {\bf 74} (2014) 12,  3134
  [arXiv:1407.2410 [hep-ex]].
  %%CITATION = ARXIV:1407.2410;%%
  %13 citations counted in INSPIRE as of 15 Apr 2015
  
  %\cite{Schael:2013ita}
\bibitem{Schael:2013ita}
  S.~Schael {\it et al.}  [ALEPH and DELPHI and L3 and OPAL and LEP Electroweak Collaborations],
  %``Electroweak Measurements in Electron-Positron Collisions at W-Boson-Pair Energies at LEP,''
  Phys.\ Rept.\  {\bf 532} (2013) 119
  [arXiv:1302.3415 [hep-ex]].
  %%CITATION = ARXIV:1302.3415;%%
  %99 citations counted in INSPIRE as of 15 Apr 2015

%\cite{Alonso:2014csa}
\bibitem{Alonso:2014csa}
  R.~Alonso, B.~Grinstein and J.~Martin Camalich,
  %``$SU(2)\times U(1)$ gauge invariance and the shape of new physics in rare $B$ decays,''
  Phys.\ Rev.\ Lett.\  {\bf 113} (2014) 241802
  [arXiv:1407.7044 [hep-ph]].
  %%CITATION = ARXIV:1407.7044;%%
  %30 citations counted in INSPIRE as of 22 May 2015
  
  
  %\cite{Hiller:2003js}
\bibitem{Hiller:2003js}
  G.~Hiller and F.~Kruger,
  %``More model independent analysis of $b \to s$ processes,''
  Phys.\ Rev.\ D {\bf 69} (2004) 074020
  [hep-ph/0310219].
  %%CITATION = HEP-PH/0310219;%%
  %145 citations counted in INSPIRE as of 31 mar 2015
  

  %\cite{Hiller:2014ula}
\bibitem{Hiller:2014ula}
  G.~Hiller and M.~Schmaltz,
  %``Diagnosing lepton-nonuniversality in $b \to s \ell \ell$,''
  JHEP {\bf 1502} (2015) 055
  [arXiv:1411.4773 [hep-ph]].
  %%CITATION = ARXIV:1411.4773;%%
  %11 citations counted in INSPIRE as of 11 Apr 2015
  


  
  
  %\cite{Blake:2015tda}
\bibitem{Blake:2015tda}
  T.~Blake, T.~Gershon and G.~Hiller,
  %``Rare b hadron decays at the LHC,''
  arXiv:1501.03309 [hep-ex].
  %%CITATION = ARXIV:1501.03309;%%
  %3 citations counted in INSPIRE as of 11 Apr 2015
  


 
  
  %\cite{Godfrey:2013eta}
\bibitem{Godfrey:2013eta}
  S.~Godfrey and T.~Martin,
  %``Z' Discovery Reach at Future Hadron Colliders: A Snowmass White Paper,''
  arXiv:1309.1688 [hep-ph].
  %%CITATION = ARXIV:1309.1688;%%
  %13 citations counted in INSPIRE as of 11 Apr 2015
  
  
    %\cite{Demir:2005ti}
\bibitem{Demir:2005ti}
  D.~A.~Demir, G.~L.~Kane and T.~T.~Wang,
  %``The Minimal U(1)' extension of the MSSM,''
  Phys.\ Rev.\ D {\bf 72} (2005) 015012
  [hep-ph/0503290].
  %%CITATION = HEP-PH/0503290;%%
  %58 citations counted in INSPIRE as of 24 Apr 2015
%
  %\cite{Hayreter:2007it}
%\bibitem{Hayreter:2007it}
  A.~Hayreter,
  %``Dilepton Signatures of Family Non-Universal U(1)-prime,''
  Phys.\ Lett.\ B {\bf 649} (2007) 191
  [hep-ph/0703269].
  %%CITATION = HEP-PH/0703269;%%
  %4 citations counted in INSPIRE as of 24 Apr 2015
%  
  %\cite{Godfrey:2008vf}
%\bibitem{Godfrey:2008vf}
  S.~Godfrey and T.~A.~W.~Martin,
  %``Identification of Extra Neutral Gauge Bosons at the LHC Using b- and t-Quarks,''
  Phys.\ Rev.\ Lett.\  {\bf 101} (2008) 151803
  [arXiv:0807.1080 [hep-ph]].
  %%CITATION = ARXIV:0807.1080;%%
  %38 citations counted in INSPIRE as of 31 Mar 2015  
%
  %\cite{Chen:2009fx}
%\bibitem{Chen:2009fx}
  M.~C.~Chen and J.~Huang,
  %``TeV Scale Seesaw and a flavorful Z-prime at the LHC,''
  Phys.\ Rev.\ D {\bf 81} (2010) 055007
  [arXiv:0910.5029 [hep-ph]].
  %%CITATION = ARXIV:0910.5029;%%
  %14 citations counted in INSPIRE as of 24 Apr 2015
%
  %\cite{Salvioni:2009jp}
%\bibitem{Salvioni:2009jp}
  E.~Salvioni, A.~Strumia, G.~Villadoro and F.~Zwirner,
  %``Non-universal minimal Z' models: present bounds and early LHC reach,''
  JHEP {\bf 1003} (2010) 010
  [arXiv:0911.1450 [hep-ph]].
  %%CITATION = ARXIV:0911.1450;%%
  %52 citations counted in INSPIRE as of 24 Apr 2015
%
  %\cite{Diener:2010sy}
%\bibitem{Diener:2010sy}
  R.~Diener, S.~Godfrey and T.~A.~W.~Martin,
  %``Unravelling an Extra Neutral Gauge Boson at the LHC using Third Generation Fermions,''
  Phys.\ Rev.\ D {\bf 83} (2011) 115008
  [arXiv:1006.2845 [hep-ph]].
  %%CITATION = ARXIV:1006.2845;%%
  %34 citations counted in INSPIRE as of 31 Mar 2015
  

 

  
  %\cite{Bouchard:2013mia}
\bibitem{Bouchard:2013mia}
  C.~Bouchard {\it et al.}  [HPQCD Collaboration],
  %``Standard Model Predictions for $B \to K \ell^+ \ell^-$ with Form Factors from Lattice QCD,''
  Phys.\ Rev.\ Lett.\  {\bf 111} (2013) 16,  162002
   [Erratum-ibid.\  {\bf 112} (2014) 14,  149902]
  [arXiv:1306.0434 [hep-ph]].
  %%CITATION = ARXIV:1306.0434;%%
  %31 citations counted in INSPIRE as of 31 Mar 2015
  
  


  
  %\cite{Bobeth:2007dw}
\bibitem{Bobeth:2007dw}
  C.~Bobeth, G.~Hiller and G.~Piranishvili,
  %``Angular distributions of anti-B ---> K anti-l l decays,''
  JHEP {\bf 0712} (2007) 040
  [arXiv:0709.4174 [hep-ph]].
  %%CITATION = ARXIV:0709.4174;%%
  %102 citations counted in INSPIRE as of 31 mar 2015
%
  %\cite{Bobeth:2011nj}
%\bibitem{Bobeth:2011nj}
  C.~Bobeth, G.~Hiller, D.~van Dyk and C.~Wacker,
  %``The Decay B --> K l^+ l^- at Low Hadronic Recoil and Model-Independent Delta B = 1 Constraints,''
  JHEP {\bf 1201} (2012) 107
  [arXiv:1111.2558 [hep-ph]].
  %%CITATION = ARXIV:1111.2558;%%
  %91 citations counted in INSPIRE as of 31 Mar 2015
  

  
  
  
%\cite{Descotes-Genon:2013vna}
\bibitem{Descotes-Genon:2013vna}
  S.~Descotes-Genon, T.~Hurth, J.~Matias and J.~Virto,
  %``Optimizing the basis of ${B} \to {K}^{*}\ell^+ \ell^-$ observables in the full kinematic range,''
  JHEP {\bf 1305} (2013) 137
  [arXiv:1303.5794 [hep-ph]].
  %%CITATION = ARXIV:1303.5794;%%
  %73 citations counted in INSPIRE as of 31 Mar 2015
        


 
  
  

  
%\cite{Beaujean:2013soa}
\bibitem{Beaujean:2013soa} 
  F.~Beaujean, C.~Bobeth and D.~van Dyk,
  %``Comprehensive Bayesian analysis of rare (semi)leptonic and radiative $B$ decays,''
  Eur.\ Phys.\ J.\ C {\bf 74}, 2897 (2014)
  [Eur.\ Phys.\ J.\ C {\bf 74}, 3179 (2014)]
  [arXiv:1310.2478 [hep-ph]].
  %%CITATION = ARXIV:1310.2478;%%
  %74 citations counted in INSPIRE as of 24 Apr 2015
  
  
  %\cite{Hurth:2013ssa}
\bibitem{Hurth:2013ssa}
  T.~Hurth and F.~Mahmoudi,
  %``On the LHCb anomaly in B $\to K^*\ell^+\ell^-$,''
  JHEP {\bf 1404} (2014) 097
  [arXiv:1312.5267 [hep-ph]].
  %%CITATION = ARXIV:1312.5267;%%
  %29 citations counted in INSPIRE as of 23 mar 2015
  


  
    %\cite{Ghosh:2014awa}
\bibitem{Ghosh:2014awa}
  D.~Ghosh, M.~Nardecchia and S.~A.~Renner,
  %``Hint of Lepton Flavour Non-Universality in $B$ Meson Decays,''
  JHEP {\bf 1412} (2014) 131
  [arXiv:1408.4097 [hep-ph]].
  %%CITATION = ARXIV:1408.4097;%%
  %21 citations counted in INSPIRE as of 23 Mar 2015

    %\cite{Hurth:2014vma}
\bibitem{Hurth:2014vma}
  T.~Hurth, F.~Mahmoudi and S.~Neshatpour,
  %``Global fits to b $\to s\ell\ell$ data and signs for lepton non-universality,''
  JHEP {\bf 1412} (2014) 053
  [arXiv:1410.4545 [hep-ph]].
  %%CITATION = ARXIV:1410.4545;%%
  %18 citations counted in INSPIRE as of 23 Mar 2015

%\cite{Altmannshofer:2015sma}
\bibitem{Altmannshofer:2015sma}
  W.~Altmannshofer and D.~M.~Straub,
  %``Implications of $b\to s$ measurements,''
  arXiv:1503.06199 [hep-ph].
  %%CITATION = ARXIV:1503.06199;%%

%\cite{Jager:2014rwa}
\bibitem{Jager:2014rwa}
  S.~Jager and J.~Martin Camalich,
  %``Reassessing the discovery potential of the $B \to K^{*} \ell^+\ell^-$ decays in the large-recoil region: SM challenges and BSM opportunities,''
  arXiv:1412.3183 [hep-ph].
  %%CITATION = ARXIV:1412.3183;%%
  %12 citations counted in INSPIRE as of 12 Apr 2015

%\cite{CMS:2014xfa}
\bibitem{CMS:2014xfa} 
  V.~Khachatryan {\it et al.}  [CMS and LHCb Collaborations],
  %``Observation of the rare $B^0_s\to\mu^+\mu^-$ decay from the combined analysis of CMS and LHCb data,''
  arXiv:1411.4413 [hep-ex].
  %%CITATION = ARXIV:1411.4413;%%
  %34 citations counted in INSPIRE as of 24 Apr 2015  

%\cite{Bobeth:2013uxa}
\bibitem{Bobeth:2013uxa} 
  C.~Bobeth {\it et al.},
  %``B_{s,d} -> l+ l- in the Standard Model with Reduced Theoretical Uncertainty,''
  Phys.\ Rev.\ Lett.\  {\bf 112}, 101801 (2014)
  [arXiv:1311.0903 [hep-ph]].
  %%CITATION = ARXIV:1311.0903;%%
  %75 citations counted in INSPIRE as of 24 Apr 2015

%\cite{Bernabeu:1986fc}
\bibitem{Bernabeu:1986fc} 
  J.~Bernabeu, G.~C.~Branco and M.~Gronau,
  %``Cp Restrictions On Quark Mass Matrices,''
  Phys.\ Lett.\ B {\bf 169}, 243 (1986).
  %%CITATION = PHLTA,B169,243;%%
  %113 citations counted in INSPIRE as of 14 Apr 2015

\end{thebibliography}
\end{document}